\documentclass[preprint,sort,compress,12pt]{elsarticle}

\usepackage{amssymb}
\usepackage{amsthm}
\usepackage{amsmath}
\usepackage{mathtools}
\usepackage{anyfontsize}
\usepackage{mathrsfs}
\usepackage{bm}
\usepackage[ruled, lined, linesnumbered, longend]{algorithm2e}
\SetKw{init}{Initialization:}
\let\oldnl\nl% Store \nl in \oldnl
\newcommand{\nonl}{\renewcommand{\nl}{\let\nl\oldnl}}% Remove line number for one line
\usepackage{array}
\usepackage{multirow}
\usepackage{listings}
\usepackage{tabu}
\usepackage{enumerate}
\usepackage{lineno}
\usepackage{fullpage}
\usepackage{xcolor}
\usepackage[colorinlistoftodos]{todonotes}
\usepackage[colorlinks=true]{hyperref}
\usepackage[capitalize,nameinlink]{cleveref}

\crefname{figure}{Fig.}{Figs.}
\Crefname{figure}{Figure}{Figures}

\usepackage[section]{placeins}
\usepackage{url}
\usepackage{textcomp}
\usepackage{gensymb}
\usepackage{soul}
\usepackage{graphicx}
\usepackage{subfig}

\usepackage{fourier} 
\usepackage{microtype}
\usepackage{array}
\usepackage{makecell}
\usepackage[export]{adjustbox}
\usepackage{float}
\usepackage{caption}
\usepackage{tikz}
\usetikzlibrary{shapes}
\biboptions{numbers,comma,round,square}
\usepackage{arydshln}
\usepackage[para,online,flushleft]{threeparttable}
\usepackage{tabularx}
\usepackage{booktabs}

\usepackage{orcidlink}

\definecolor{myred}{RGB}{214,39,40}
\definecolor{mygray}{RGB}{176,176,176}
\definecolor{myorange}{RGB}{255,127,14}
\definecolor{mygreen}{RGB}{44,160,44}
\definecolor{mylightgray}{RGB}{204,204,204}
\definecolor{mypurple}{RGB}{148,103,189}
\definecolor{mybrown}{RGB}{140,86,75}
\definecolor{steelblue}{RGB}{31,119,180}
\definecolor{intraray}{RGB}{127,193,219}
\definecolor{hemitrichs}{RGB}{26,74,93}

\theoremstyle{definition}

\theoremstyle{remark}

\graphicspath{ {./figs/} }

\DontPrintSemicolon
\definecolor{orcidlogocol}{HTML}{A6CE39}
\begin{document}
\makeatletter
\def\ps@pprintTitle{%
  \let\@oddhead\@empty
  \let\@evenhead\@empty
  \let\@oddfoot\@empty
  \let\@evenfoot\@oddfoot
}
\makeatother
\begin{frontmatter}

\title{Retrainable physics-integrated neural differentiable modeling of sintering across material systems}

\author[ndCBE]{Zeping Chen}
\author[ndNuclear,ndChem]{Ani Aprahamian}
\author[ndNuclear]{Khachatur V. Manukyan}
\author[ndCBE,ndAME]{Tengfei Luo\texorpdfstring{\corref{corjw}}{}}

\address[ndCBE]{Department of Chemical and Biomolecular Engineering, University of Notre Dame, Notre Dame, IN, USA}
\address[ndAME]{Department of Aerospace and Mechanical Engineering, University of Notre Dame, Notre Dame, IN, USA}
\address[ndNuclear]{Nuclear Science Laboratory, Department of Physics \& Astronomy, University of Notre Dame, Notre Dame, IN, USA}
\address[ndChem]{Department of Chemistry \& Biochemistry, University of Notre Dame, Notre Dame, IN, USA}

\cortext[corjw]{Corresponding author: tluo@nd.edu}

\begin{abstract}
Sintering is widely used to manufacture ceramics for structural, electronic, and energy applications, but coupled densification and grain growth, material-dependent kinetics, and sparse measurements complicate predictive modeling and process design. Without reliable simulation, processing schedules are commonly developed through repeated furnace trials and post-process characterization. Here, we present Sinter-PiNDiff, a retrainable sintering physics-integrated neural differentiable framework for predicting density and grain-size evolution across ceramic systems. Two neural networks learn material-dependent densification and grain-growth coefficients within coupled sintering rate equations, while a smooth density-saturation factor attenuates densification as theoretical density is approached. The common governing structure, network architecture, and training procedure were fitted independently to published measurements for MgO, Al-doped ZnO, and CaO-doped ThO$_2$. Condition-wise held-out tests evaluate predictions at unseen temperatures and compositions. Sinter-PiNDiff achieved the lowest mean error in all twelve material--metric comparisons with multilayer-perceptron and residual-network baselines. For MgO, Al-doped ZnO, and CaO-doped ThO$_2$, respectively, density normalized root-mean-square errors were $14.6\%$, $10.8\%$, and $14.4\%$, and grain-size normalized root-mean-square errors were $8.6\%$, $12.1\%$, and $19.3\%$. Removing the evolving density from both neural-network inputs increased both density and grain-size trajectory errors in all three systems and increased ten of the twelve aggregate errors examined, supporting density-dependent kinetic feedback. Deep ensembles estimated model disagreement, although empirical coverage showed that the resulting bands were not calibrated and did not capture all model--data discrepancies. These results establish Sinter-PiNDiff as a retrainable framework for sparse-data prediction and uncertainty-informed selection of sintering conditions.
\end{abstract}

\end{frontmatter}

\newpage

%%%%%%%%%%%%%%%%%%%%%%%%%%%%%%%%%%%%%%%%%%%%%%%%%%%%%%%%%%%%%%%%%%%%%%%%%%%%%%%%%%%%%%%%%%%%%%%%%%%%%%
%%%%%%%%%%%%%%%%%%%%%%%%%%%%%%%%%%%%%%%%%%%%%%%%%%%%%%%%%%%%%%%%%%%%%%%%%%%%%%%%%%%%%%%%%%%%%%%%%%%%%%

\section{Introduction}

Sintering is a central process in ceramic manufacturing, enabling porous powder compacts to be consolidated into polycrystalline components for structural, electronic, and energy applications~\cite{German1996,Bordia2017}. Through thermally activated mass transport, it determines the development of relative density and grain size, which strongly influence the mechanical, thermal, and functional properties of the final material~\cite{Olevsky1998,Bordia2017}. Classical treatments distinguish the initial, intermediate, and final stages of sintering and relate neck growth, shrinkage, and coarsening to capillarity-driven transport along surfaces, grain boundaries, and crystal lattices~\cite{Herring1950Scale,Kingery1955InitialStages,Johnson1969DiffusionCoefficients,Rahaman2003CeramicProcessing,Kang2005Sintering}. A central challenge is balancing densification against grain growth: pore elimination increases density, whereas excessive coarsening and pore--grain-boundary interactions can isolate residual pores and hinder further densification~\cite{Brook1969PoreGrain,Hillert1965GrainGrowth,Atkinson1988GrainGrowth}. Although two-step sintering can partly separate densification from final-stage grain growth, the accessible kinetic window remains material dependent~\cite{Chen2000Nanocrystalline}. At the particle scale, sintering proceeds through capillarity-driven mass transport along several pathways, including surface, grain-boundary, and lattice diffusion. These pathways contribute differently to microstructural evolution: surface diffusion primarily promotes neck growth and coarsening without substantial densification, whereas grain-boundary and lattice diffusion can drive particle rearrangement, pore shrinkage, and bulk densification~\cite{Herring1950Scale,Kingery1955InitialStages,Johnson1969DiffusionCoefficients,Rahaman2003CeramicProcessing,Kang2005Sintering,Karacasulu2025}. Because these mechanisms have different activation energies and are affected differently by the evolving pore structure and dopant chemistry, their relative contributions depend on temperature, hold time, and composition. These coupled effects complicate the selection of processing conditions. Without predictive simulation, process development relies heavily on repeated furnace cycles followed by density measurements and microstructural characterization, making broad exploration of temperature, hold time, and composition experimentally costly~\cite{Shi2023}. Physics-based simulations offer a means to reduce this burden by predicting the evolution of density and grain size, clarifying their coupled kinetics, and screening candidate processing conditions before fabrication~\cite{Ashby1974SinteringDiagrams,Olevsky1998,Bordia2017}. Extending these capabilities across ceramic systems, however, requires models that can accommodate material-dependent kinetics despite sparse calibration data.

Conventional sintering models provide a physical basis for these predictions. Mechanistic descriptions span reduced-order scaling relations and sintering diagrams~\cite{Ashby1974SinteringDiagrams}, continuum models that relate densification to capillary driving forces and the evolving compact structure~\cite{Olevsky1998,Bordia2017}, and phase-field, mesoscale, and variational formulations that resolve pore evolution, grain-boundary motion, and particle-scale mass transport~\cite{Wang2006PhaseField,Tikare2010Mesoscale,Wakai2011Mechanics}. Their predictive application, however, requires more than assigning values to a fixed set of material parameters; it also requires identifying which physical and chemical mechanisms are active for a particular powder, composition, atmosphere, and thermal history. Depending on the material and processing route, sintering may involve surface, lattice, and grain-boundary diffusion, together with vapor transport, viscous or plastic flow, particle rearrangement, and grain-boundary sliding~\cite{Ashby1974SinteringDiagrams,Bordia2017}. These mechanisms contribute differently to neck growth, shrinkage, pore elimination, and coarsening. Their relative importance is governed not only by diffusivities and activation energies, but also by particle size and morphology, green density and pore topology, defect chemistry, surface and grain-boundary chemistry, interfacial energies, gas atmosphere, and applied pressure. Composition is therefore more than an additional scalar process variable. In ionic ceramics, dopants can modify point-defect populations and charge-compensation pathways, segregate to surfaces or grain boundaries, alter grain-boundary mobility, or react to form secondary phases. For example, Al additions to ZnO affect both densification and grain growth, with ZnAl$_2$O$_4$ forming as a secondary phase at ZnO grain boundaries and inhibiting their migration~\cite{Han2001}. Furthermore, when only limited density and grain-size measurements are available, multiple combinations of effective parameters may fit the observations similarly while representing different underlying kinetics~\cite{Saleem2026}. Thus, although different ceramics share broad sintering mechanisms, their effective kinetic relations reflect material-specific chemistry and microstructure and cannot generally be represented by one fixed set of parameters.

Machine learning (ML) offers a complementary means of representing these complex relationships from data and now supports materials-property prediction, discovery, and inverse design across a broad range of chemistries and length scales~\cite{Butler2018MaterialsML,Schmidt2019MaterialsML,Batra2021MaterialsIntelligence}. Deep neural networks provide flexible nonlinear approximations and have been widely applied to materials-property prediction and microstructural analysis~\cite{Choudhary2022}, including the characterization, reconstruction, and structure--property mapping of heterogeneous microstructures~\cite{Bostanabad2018Microstructure,Cecen2018StructureProperty}. Data-driven models have also been used to predict sintered microstructures from processing parameters~\cite{Tang2021}. However, learning density and grain-size evolution directly from measurements places substantial demands on the amount and coverage of the training data. These demands are particularly restrictive for sintering datasets, in which complete temporal histories may be available for only a few temperatures or compositions. Without explicit physical structure, a model must infer both the material response and its temporal evolution from these sparse observations, leaving predictions at unseen conditions weakly constrained.

Physics-informed ML addresses this challenge by incorporating scientific knowledge through the inputs, loss function, model architecture, or a hybrid combination of learned and mechanistic components~\cite{Willard2022ScientificKnowledge,Karniadakis2021,chen_PIDDM,CHEN_composite}. Physics-informed neural networks, for example, supplement measurement-based training with governing-equation residuals~\cite{Raissi2019}. Neural ordinary differential equations and universal differential equations instead learn unknown dynamical terms while retaining numerical time integration within the trainable model~\cite{Chen2018NeuralODE,Rackauckas2020UDE}. In sintering, related approaches have introduced neural constitutive laws for finite-element analysis, learned deformation behavior from densification measurements, and combined physical models with deep learning for metal additive manufacturing~\cite{He2024,Saleem2026,Kassab2025,Bai2026}. These studies demonstrate the value of combining experimental information with physical constraints. A complementary challenge for prediction is to describe density and grain-size histories together using a formulation that can be retrained across chemically distinct ceramics. Such a formulation must retain the shared structure of sintering evolution while accommodating material-dependent kinetics and incomplete observations of the evolving microstructure.

Physics-integrated neural differentiable modeling (PiNDiff) provides a basis for addressing this challenge. Originally developed for composite curing, PiNDiff embeds neural representations of unresolved physical terms within partially known governing equations~\cite{AKHARE_curing}. The neural components and numerical time integration form a single differentiable computational model, allowing discrepancies between predicted and measured process histories to train the unknown terms. The framework was subsequently extended to probabilistic modeling of isothermal chemical vapor infiltration using limited experimental data~\cite{Akhare2024}. This separation between retained evolution equations and learned kinetic relations is well suited to sintering, where the dependence of densification on grain size provides useful physical structure, but effective transport and growth coefficients vary with material and processing conditions. Applying this strategy across independent ceramic systems offers a way to examine whether a common model structure can capture these differences without prescribing a separate empirical formulation for each system.

\begin{figure}[ht]
    \centering
    \includegraphics[width=\textwidth]{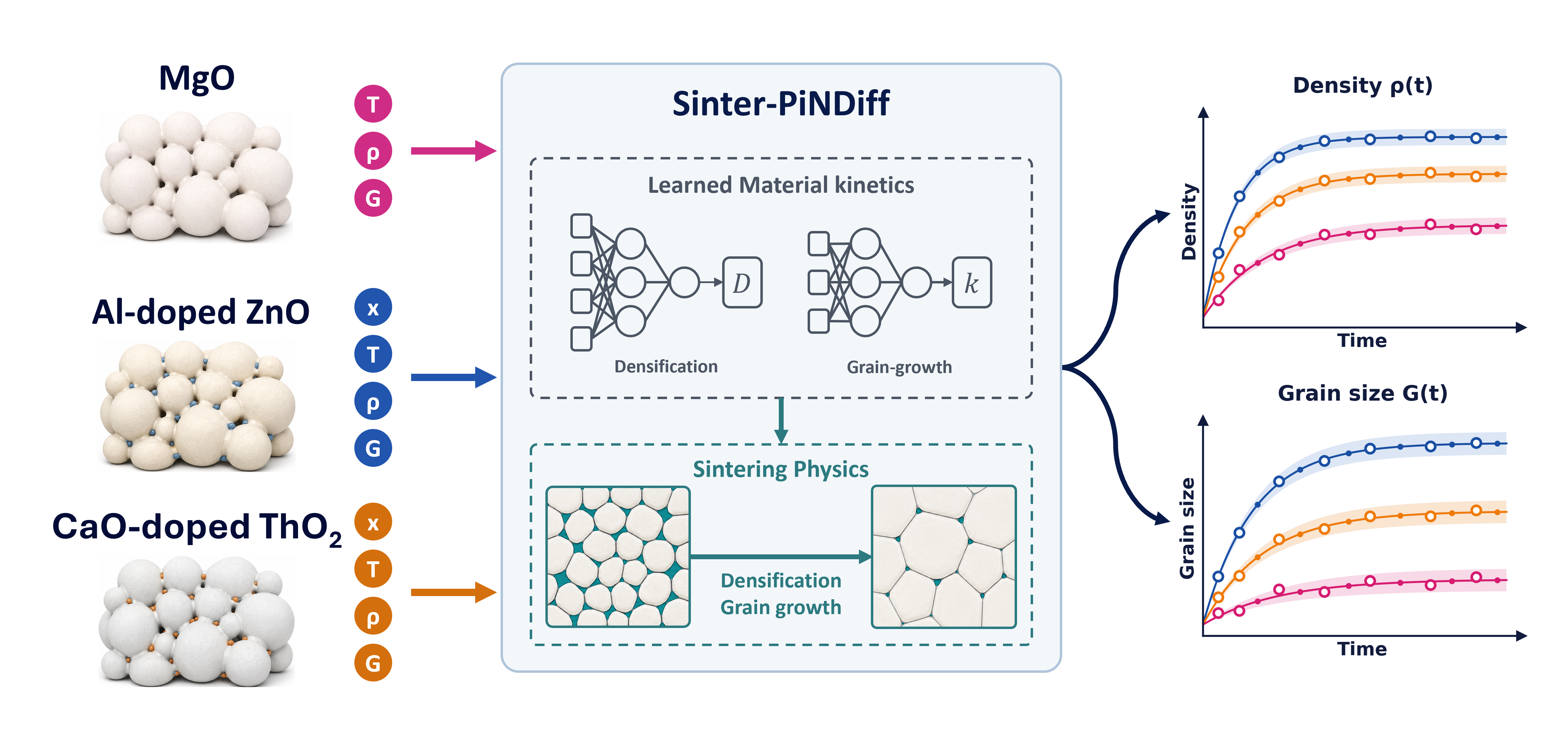}
    \caption{\textbf{Overview of the Sinter-PiNDiff framework.} Neural networks learn material-dependent densification and grain-growth coefficients, which are embedded within sintering evolution equations to predict density and grain-size histories. The common governing structure and network architecture are independently retrained for MgO, Al-doped ZnO, and CaO-doped ThO$_2$, with separate weights learned for each material system. Schematic output curves and shaded bands illustrate the predicted evolution and ensemble uncertainty, respectively.}
    \label{fig:sinter-pindiff-overview}
\end{figure}

Here, we develop Sinter-PiNDiff, a retrainable sintering physics-integrated neural differentiable framework for material-specific prediction across different material systems (\cref{fig:sinter-pindiff-overview}). Neural networks learn effective densification and grain-growth coefficients within a common set of rate equations. The governing structure, network architecture, and training procedure are retained across material systems, with separate network weights learned for each system. A deep ensemble provides an estimate of model uncertainty~\cite{Lakshminarayanan2017}; with appropriate validation, such estimates can support adaptive sampling and closed-loop materials experimentation~\cite{Lookman2019ActiveLearning,Kusne2020ClosedLoop}.

We evaluate the framework using experimental measurements for MgO, Al-doped ZnO, and CaO-doped ThO$_2$ reported in previously published studies~\cite{Gupta1971,Han2001,Laha1971}. These datasets were selected because they collectively provide the observations needed to evaluate coupled densification and grain growth under sparse and nonuniform experimental coverage. All three contain relative-density measurements at multiple isothermal temperatures and at least some grain-size measurements, while the ZnO and ThO$_2$ datasets additionally span multiple dopant concentrations. MgO therefore tests temperature-dependent prediction in a single-composition system, whereas Al-doped ZnO and CaO-doped ThO$_2$ also permit evaluation of composition-dependent behavior.

Beyond data availability, the three systems were selected to span distinct sintering environments arising from differences in crystal structure, bonding, and defect chemistry. Sintering is driven by the reduction of interfacial energy, while its kinetics depend on atomic diffusion; bonding therefore affects densification and grain growth by altering defect formation and migration barriers~\cite{Bordia2017,Carrasco2004}. MgO represents the strongly ionic limit, with approximately 80\% ionic character and a rock-salt structure, whereas wurtzite ZnO has mixed ionic--covalent bonding with a substantial covalent contribution. Fluorite ThO$_2$ is predominantly ionic but also exhibits measurable Th--O covalency~\cite{Phillips1970,Boyer1983MgO,Jaffe1993ZnO,Wang_2010}. These differences produce distinct diffusion pathways and temperature sensitivities: MgO provides an undoped, highly ionic lattice-diffusion case~\cite{Gupta1971}; Ca$^{2+}$ substitution in ThO$_2$ creates oxygen vacancies that influence cation migration and reduce the apparent activation energy~\cite{Laha1971}; and Al addition to ZnO alters defect chemistry while forming ZnAl$_2$O$_4$ particles that inhibit densification and pin grain boundaries~\cite{Han2001}. The datasets therefore cover undoped, vacancy-modified, and second-phase-affected sintering and test whether the shared rate-equation structure can accommodate this physical range after material-specific retraining.

Some temperature--composition trajectories were withheld from training for condition-wise evaluation. The resulting splits test temporal rollout at unseen processing conditions, temperature interpolation across all three systems, composition interpolation in Al-doped ZnO and CaO-doped ThO$_2$, and composition extrapolation to 8.80 mol\% CaO. Comparisons with multilayer-perceptron and residual-network baselines show lower mean errors for all reported dynamic and final-state metrics across the three systems, supporting the utility of the shared formulation within the investigated material and processing ranges.

The following sections present the governing equations, neural-network architecture, training procedure, and experimental datasets. We then examine predictive accuracy and uncertainty across the three material systems and compare performance against data-driven baselines. Finally, we discuss the framework's applicability, current limitations, and opportunities for extending it to experimental design and process optimization.

%%%%%%%%%%%%%%%%%%%%%%%%%%%%%%%%%%%%%%%%%%%%%%%%%%%%%%%%%%%%%%%%%%%%%%%%%%%%%%%%%%%%%%%%%%%%%%%%%%%%%%
%%%%%%%%%%%%%%%%%%%%%%%%%%%%%%%%%%%%%%%%%%%%%%%%%%%%%%%%%%%%%%%%%%%%%%%%%%%%%%%%%%%%%%%%%%%%%%%%%%%%%%

\section{Methodology}

\subsection{Physical basis of coupled densification and grain growth}

Pressureless solid-state sintering is driven by the reduction of interfacial free energy through pore elimination and microstructural coarsening. Densifying diffusion paths transport matter into interparticle contacts and reduce pore volume, whereas surface transport can enlarge particle necks without producing equivalent macroscopic shrinkage~\cite{Herring1950Scale,Kingery1955InitialStages,Johnson1969DiffusionCoefficients}. Grain-boundary migration simultaneously reduces interfacial area and increases the mean grain size $G$, while pore--boundary interactions can retard or modify this coarsening process~\cite{Brook1969PoreGrain,Wakai2011Mechanics}. The evolving grain and pore structures alter diffusion distances, capillary driving forces, and the available transport paths, thereby coupling grain growth to the densification rate~\cite{Coble1961a,Rahaman2003CeramicProcessing,Kang2005Sintering}.

In the classical sintering relations summarized by Karacasulu et al.~\cite{Karacasulu2025}, densification and grain growth are described by
\begin{align}
    \dot{\rho}
    &= \frac{C_{0}D_{0}\alpha\Gamma(\rho)}{G^{m}RT}
    \exp\!\left(-\frac{Q_{\rho}}{RT}\right),
    \label{eq:classical-densification}\\
    \dot{G}
    &= \frac{k_{0}}{G^{p}}
    \exp\!\left(-\frac{Q_{G}}{RT}\right).
    \label{eq:classical-grain-growth}
\end{align}
Here, $\rho$ is the relative density, $\alpha$ is the surface energy, $\Gamma(\rho)$ is a stress-intensification factor representing the evolving pore geometry, and $C_{0}$ is the fixed geometric and transport prefactor. The quantities $k_{0}$, $Q_{\rho}$, and $Q_{G}$ are the grain-growth prefactor and the activation energies for densification and grain growth, respectively. The absolute temperature is $T$, and $R$ is the ideal-gas constant. Material and composition dependence of the kinetic parameters is left implicit in these expressions.

In this work, $m=3$ and $p=2$ are fixed for all material systems. The choice $m=3$ follows the volume-diffusion scaling of classical sintering models~\cite{Coble1961a,Karacasulu2025}, while $p=2$ adopts the grain-growth form associated with bulk mass transport~\cite{Karacasulu2025}. For constant temperature and kinetic parameters, the latter gives $G^{3}(t)-G^{3}(t_{0})=3k_{0}\exp[-Q_{G}/(RT)](t-t_{0})$, consistent with the cubic grain-growth behavior reported during sintering by Coble~\cite{Coble1961b}. Classical scaling analyses and sintering diagrams show that apparent power-law dependences vary with the transport path, geometry, and sintering stage~\cite{Herring1950Scale,Johnson1969DiffusionCoefficients,Ashby1974SinteringDiagrams}. The fixed exponents are therefore reduced-order kinetic assumptions; their use does not establish the dominant microscopic transport mechanism in each experimental dataset.

To obtain the rate equations used by Sinter-PiNDiff, the unresolved kinetic factors are grouped into two effective coefficients. The effective densification and grain-growth coefficients are then defined as
\begin{align}
    D_{nn} &\equiv D_{0}\alpha\operatorname{\Gamma}(\rho)
    \exp\!\left(-\frac{Q_{\rho}}{RT}\right),
    \label{eq:effective-densification-coefficient}\\
    k_{nn} &\equiv k_{0}
    \exp\!\left(-\frac{Q_{G}}{RT}\right).
    \label{eq:effective-grain-growth-coefficient}
\end{align}
Substitution into \cref{eq:classical-densification,eq:classical-grain-growth} gives the retained physical backbone,
\begin{align}
    \psi(\rho)
    &=
    \tanh\!\left(
    \frac{1-\rho}{\varepsilon_{\rho}}
    \right),
    \qquad
    \varepsilon_{\rho}=0.05,
    \label{eq:density-saturation-factor}\\
    \dot{\rho}
    &=
    \frac{C_{0}\psi(\rho)D_{\mathrm{nn}}}
    {G^{m}RT},
    \qquad m=3,
    \label{eq:densification-rate}\\
    \dot{G}
    &=
    \frac{k_{\mathrm{nn}}}{G^{p}},
    \qquad p=2.
    \label{eq:grain-growth-rate}
\end{align}
This reformulation preserves the explicit grain-size dependence while describing the coupled evolution of $\bm{v}^{(s)}(t)=[\rho(t),G(t)]^{\mathsf{T}}$ for a material system $s$ under prescribed isothermal conditions $\bm{u}=[T,\chi_{\mathrm{d}},\chi_{\mathrm{b}}]^{\mathsf{T}}$. Here, $\chi_{\mathrm{d}}$ and $\chi_{\mathrm{b}}$ denote the molar fractions of the dopant and base oxide, respectively, with $\chi_{\mathrm{d}}=0$ and $\chi_{\mathrm{b}}=1$ for MgO. Unresolved thermal and microstructural contributions are represented through learned kinetic coefficients: the grain-growth network learns $k_{0}\exp[-Q_{G}/(RT)]$, while the densification network learns the combined contribution
$D_0\alpha\Gamma(\rho)\exp[-Q_{\rho}/(RT)]$. These composite quantities are learned directly, without separately identifying the constituent activation energies, Arrhenius prefactors, surface energy, or stress-intensification factor.

%%%%%%%%%%%%%%%%%%%%%%%%%%%%%%%%%%%%%%%%%%%%%%%%%%%%%%%%%%%%%%%%%%%%%%%%%%%%%%%%%%%%%%%%%%%%%%%%%%%%%%

\subsection{Sinter-PiNDiff architecture}

For each material system, two fully connected neural networks approximate the effective densification and grain-growth coefficients. Both networks receive the prescribed temperature and composition together with the current density. At numerical step $n$, their outputs are
\begin{align}
    D_{nn} &= \mathcal{N}_{\rho}\!\left(
    T,\rho_n,\chi_{\mathrm{d}},\chi_{\mathrm{b}}
    \right),
    \label{eq:density-closure}\\
    k_{nn} &= \mathcal{N}_{G}\!\left(
    T,\rho_n,\chi_{\mathrm{d}},\chi_{\mathrm{b}}
    \right).
    \label{eq:grain-closure}
\end{align}
Although temperature and composition remain fixed during each isothermal hold, both $D_{nn}$ and $k_{nn}$ can vary as the density evolves. Dependence on grain size is retained explicitly through $G^{-m}$ in the densification equation and $G^{-p}$ in the grain-growth equation. This structure couples the two processes through the influence of grain size on densification and the influence of density on the learned kinetic coefficients.

\begin{figure}[ht]
    \centering
    \includegraphics[width=0.9\textwidth]{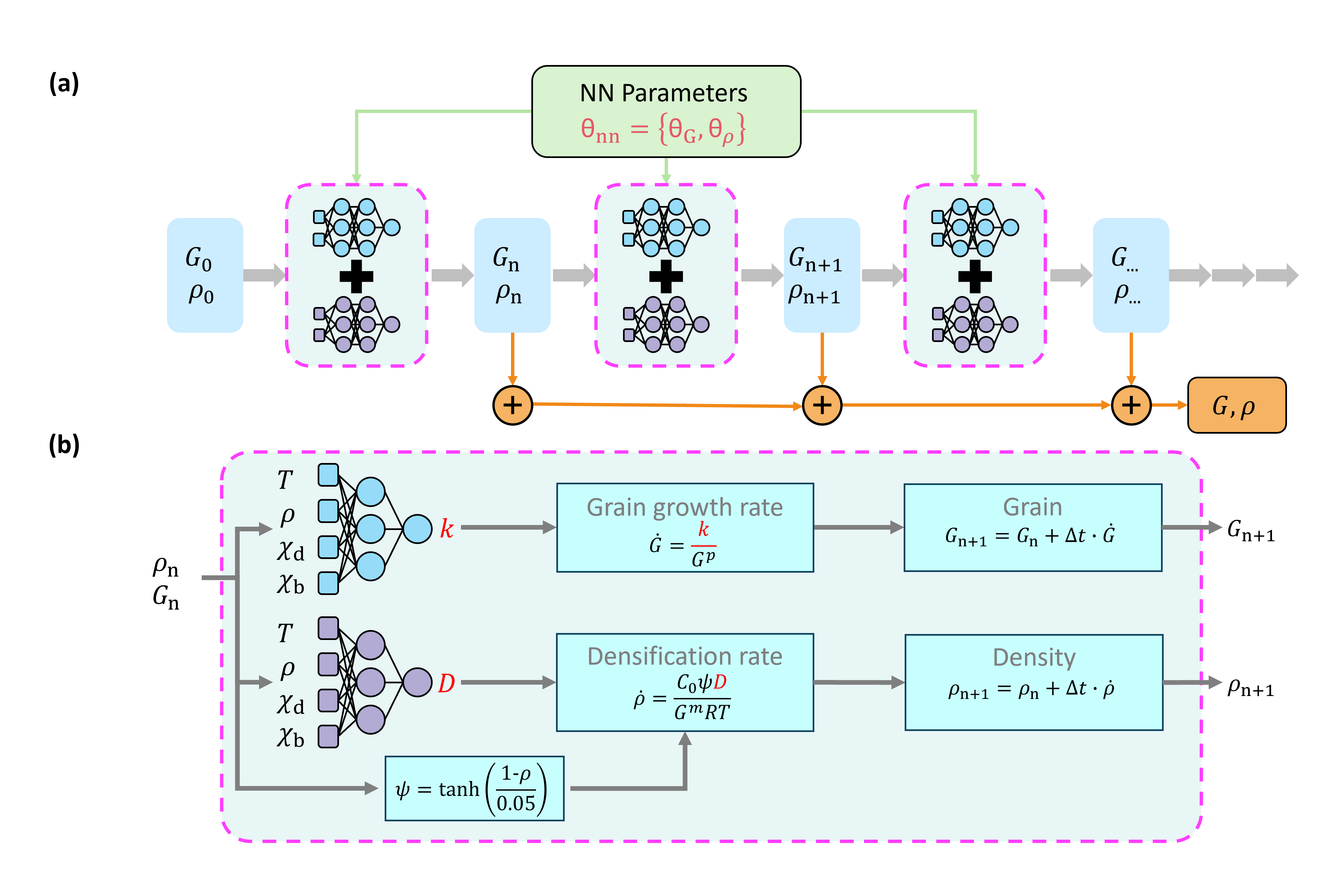}
    \caption{\textbf{Differentiable autoregressive architecture of Sinter-PiNDiff.} \textbf{(a)} Density and grain size evolve through recurrent Sinter-PiNDiff cells with neural-network parameters shared across time steps. Predictions at observation times contribute to the training objective, and gradients propagate through the complete rollout. \textbf{(b)} Within each cell, two neural networks receive the temperature $T$, current density $\rho_n$, and composition $(\chi_{\mathrm{d}},\chi_{\mathrm{b}})$ to predict the effective coefficients $k_{\mathrm{nn},n}$ and $D_{\mathrm{nn},n}$. The density-saturation factor $\psi(\rho_n)$ attenuates the densification rate as the relative density approaches one. The resulting rates are integrated using an explicit Euler update to obtain $G_{n+1}$ and $\rho_{n+1}$.}
    \label{fig:sinter-pindiff-architecture}
\end{figure}

The neural network and rate equations form an autoregressive operator, as shown in \cref{fig:sinter-pindiff-architecture}. Given the state $\bm{v}_n=[\rho_n,G_n]^{\mathsf{T}}$ at time $t_n$, both coefficients are evaluated using the current density $\rho_n$. Their values are then inserted into the retained rate laws, and the state is advanced using an explicit Euler update,
\begin{align}
    G_{n+1}
    &=
    G_n+\Delta t_n
    \frac{k_{\mathrm{nn},n}}{G_n^{p}},
    \label{eq:grain-update}\\
    \rho_{n+1}
    &=
    \rho_n+\Delta t_n
    \frac{C_0\psi(\rho_n)D_{\mathrm{nn},n}}
    {G_n^{m}RT}.
    \label{eq:density-update}
\end{align}
where $\Delta t_n=t_{n+1}-t_n$, with $m=3$ and $p=2$. The updated state is passed to the next step, where both networks are evaluated using $\rho_{n+1}$ and the rate equations use $G_{n+1}$.

This construction is related to neural and universal differential-equation formulations in which unknown rate components are represented by neural networks and trained through a numerical time integrator~\cite{Chen2018NeuralODE,Rackauckas2020UDE}. Automatic differentiation propagates sensitivities through the recurrent rollout, allowing observations at later times to update the kinetic closures evaluated at preceding steps~\cite{Baydin2018AutoDiff}.

\subsubsection{End-to-end autoregressive training}

The recurrent solver was implemented as a single differentiable program so that discrepancies at any observed time could update the kinetic closures responsible for the complete preceding trajectory. For a training case $c$, the model is initialized with $\bm{v}_{0}^{(c)}$ and rolled forward over the full isothermal hold. A binary observation mask $M_{q,n}^{(c)}$ indicates whether quantity $q\in\{\rho,G\}$ was reported at time $t_n$, permitting joint training with different levels of density and grain-size supervision. The data loss is a sum of normalized mean squared errors,
\begin{equation}
    \begin{aligned}
        \mathcal{L}_{\mathrm{data}}(\bm{\theta})
        ={}&\sum_{q\in\{\rho,G\}}\frac{w_q}{N_q}
        \sum_{c\in\mathcal{C}_{\mathrm{train}}}\sum_n
        M_{q,n}^{(c)}\\
        &\times\left(
        \widehat{q}_{n}^{(c)}(\bm{\theta})-q_{n}^{(c)}
        \right)^2,
    \end{aligned}
    \label{eq:data-loss}
\end{equation}
where $N_q=\sum_{c\in\mathcal{C}_{\mathrm{train}}}\sum_n M_{q,n}^{(c)}$ is the number of available training observations of $q$. The normalization weight is the inverse of the mean squared observations,
\begin{equation}
    w_q=
    \left[
        \frac{1}{N_q}
        \sum_{c\in\mathcal{C}_{\mathrm{train}}}\sum_n
        M_{q,n}^{(c)}\left(q_n^{(c)}\right)^2
    \right]^{-1}.
    \label{eq:loss-normalization}
\end{equation}
Each contribution therefore measures the mean squared prediction error relative to the mean squared magnitude of the corresponding observations, making the loss dimensionless and invariant to multiplicative rescaling of either property. The parameter set $\bm{\theta}=\{\bm{\theta}_{\rho},\bm{\theta}_{G}\}$ contains the trainable weights and biases of both neural networks. The total objective combines \cref{eq:data-loss} with weight regularization,
\begin{equation}
    \mathcal{L}(\bm{\theta})
    =\mathcal{L}_{\mathrm{data}}(\bm{\theta})
    +\lambda\lVert\bm{\theta}\rVert_2^2.
    \label{eq:total-loss}
\end{equation}
Because the governing rate equations are embedded directly in the recurrent update, no separate collocation-point residual is required to impose \cref{eq:densification-rate,eq:grain-growth-rate}. Automatic differentiation through the complete rollout provides $\partial\mathcal{L}/\partial\bm{\theta}$, and all neural-network parameters are optimized simultaneously. The train--test assignment is fixed before preprocessing. Numerical scaling factors are prescribed for each dataset according to the physical orders of magnitude of the relevant quantities.

\subsubsection{Deep-ensemble uncertainty quantification}

Incomplete kinetic knowledge and sparse experimental coverage introduce epistemic uncertainty into the learned closures. Following the deep-ensemble approach used in probabilistic PiNDiff~\cite{Akhare2024}, this uncertainty is approximated using $N_{\mathrm{e}}=5$ independently initialized Sinter-PiNDiff models. Every member uses the same architecture, governing equations, training data, and optimization settings, with a different random initialization of the neural-network weights. For an output $q\in\{\rho,G\}$, the ensemble mean and variance at condition $\bm{u}$ and time $t$ are
\begin{align}
    \overline{q}(t,\bm{u})
    &=\frac{1}{N_{\mathrm{e}}}
    \sum_{j=1}^{N_{\mathrm{e}}}
    \widehat{q}^{(j)}(t,\bm{u}),
    \label{eq:ensemble-mean}\\
    \sigma_q^2(t,\bm{u})
    &=\frac{1}{N_{\mathrm{e}}}
    \sum_{j=1}^{N_{\mathrm{e}}}
    \left[
    \widehat{q}^{(j)}(t,\bm{u})-\overline{q}(t,\bm{u})
    \right]^2.
    \label{eq:ensemble-var}
\end{align}
The reported prediction is $\overline{q}$, and the shaded uncertainty region is $\overline{q}\pm3\sigma_q$. Since the ensemble members do not predict a separate observation-noise variance, this band represents variability among the learned models associated with finite data, parameter non-uniqueness, and extrapolation from the training conditions. Deep ensembles provide a practical approximation of epistemic uncertainty, but their spread is not automatically a calibrated prediction interval~\cite{Gawlikowski2023UQ}. Predictive uncertainty can degrade under dataset shift~\cite{Ovadia2019DatasetShift}, and neural-network confidence generally requires explicit calibration and empirical coverage assessment~\cite{Guo2017Calibration}. The reported $\pm3\sigma_q$ bands are therefore interpreted as descriptive disagreement among independently trained models; they do not provide a complete estimate of experimental uncertainty or intervals with a prescribed coverage probability.

%%%%%%%%%%%%%%%%%%%%%%%%%%%%%%%%%%%%%%%%%%%%%%%%%%%%%%%%%%%%%%%%%%%%%%%%%%%%%%%%%%%%%%%%%%%%%%%%%%%%%%%%%%%%%%%%%%%%%%%%%%%%%%%%%%%%%%%%
%%%%%%%%%%%%%%%%%%%%%%%%%%%%%%%%%%%%%%%%%%%%%%%%%%%%%%%%%%%%%%%%%%%%%%%%%%%%%%%%%%%%%%%%%%%%%%%%%%%%%%%%%%%%%%%%%%%%%%%%%%%%%%%%%%%%%%%%
\subsection{Experimental datasets and condition-wise train--test design}
\label{sec:experimental_data}

\subsubsection{Experimental datasets and initial conditions}

The framework was evaluated using data from three independent, previously published studies: MgO reported by Gupta~\cite{Gupta1971}, Al-doped ZnO reported by Han et al.~\cite{Han2001}, and pure and CaO-doped $\mathrm{ThO_2}$ reported by Laha and Das~\cite{Laha1971}. Each unique temperature--composition pair was treated as one experimental case containing all reported times and observables.

For MgO, Gupta reported an approximate particle size of $0.2~\mu\mathrm{m}$ for the calcined powder before compaction and presintering~\cite{Gupta1971}. The powder compacts were subsequently presintered at $900~^{\circ}\mathrm{C}$ for $1$~h, after which specimens with densities of $1.66$--$1.67~\mathrm{g\,cm^{-3}}$, corresponding to approximately $47\%$ of theoretical density, were selected for the isothermal experiments. Because the grain size immediately after presintering was not reported directly, the value $G_0=2.0~\mu\mathrm{m}$ used by Sinter-PiNDiff is treated as an effective model initialization value rather than as the powder-particle size reported by Gupta. The experiments span $1450$--$1650~^{\circ}\mathrm{C}$ in $50~^{\circ}\mathrm{C}$ increments. The maximum reported dwell time is $1000$~min, although the available duration depends on temperature and observable. Density and grain-size histories are available from $1450$ to $1600~^{\circ}\mathrm{C}$, whereas only density was reported at $1650~^{\circ}\mathrm{C}$.

The Al-doped ZnO dataset contains Al concentrations of $0$, $0.08$, $0.32$, $0.60$, and $1.20$~mol\% over temperatures from $1100$ to $1400~^{\circ}\mathrm{C}$. Han et al.~reported a mean starting powder-particle size of $0.26~\mu\mathrm{m}$, which is used as $G_0$, and a green density of approximately $60\%$ of theoretical density~\cite{Han2001}. Time-resolved density and grain-size measurements extending to $480$~min are available at $1200~^{\circ}\mathrm{C}$, whereas the remaining temperatures provide density measurements after a $120$-min dwell.

The $\mathrm{ThO_2}$--CaO dataset contains CaO concentrations of $0$, $2.31$, $4.25$, and $8.80$~mol\% at $1300$, $1400$, $1500$, $1550$, and $1600~^{\circ}\mathrm{C}$. A value of $G_0=2.0~\mu\mathrm{m}$ is used to initialize all compositions, and the representative initial density is approximately $66\%$ of theoretical density~\cite{Laha1971}. The reported dwell times are $5$, $10$, $20$, $30$, $60$, and $120$~min. Density measurements are available at all five temperatures, whereas grain-size measurements are available at $1500$, $1550$, and $1600~^{\circ}\mathrm{C}$. Because the initial densities vary with composition, the condition-specific values, rather than the rounded representative value reported above, are used to initialize the individual model rollouts.

%%%%%%%%%%%%%%%%%%%%%%%%%%%%%%%%%%%%%%%%%%%%%%%%%%%%%%%%%%%%%%%%%%%%%%%%%%%%%%%%%%%%%%%%%%%%%%%%%%%%%%%%%%%%%%%%%%%%%%%%%%%%%%%%%%%%%%%%

\subsubsection{Condition-wise train--test split}

\begin{figure}[ht]
    \centering
    \includegraphics[width=\textwidth]{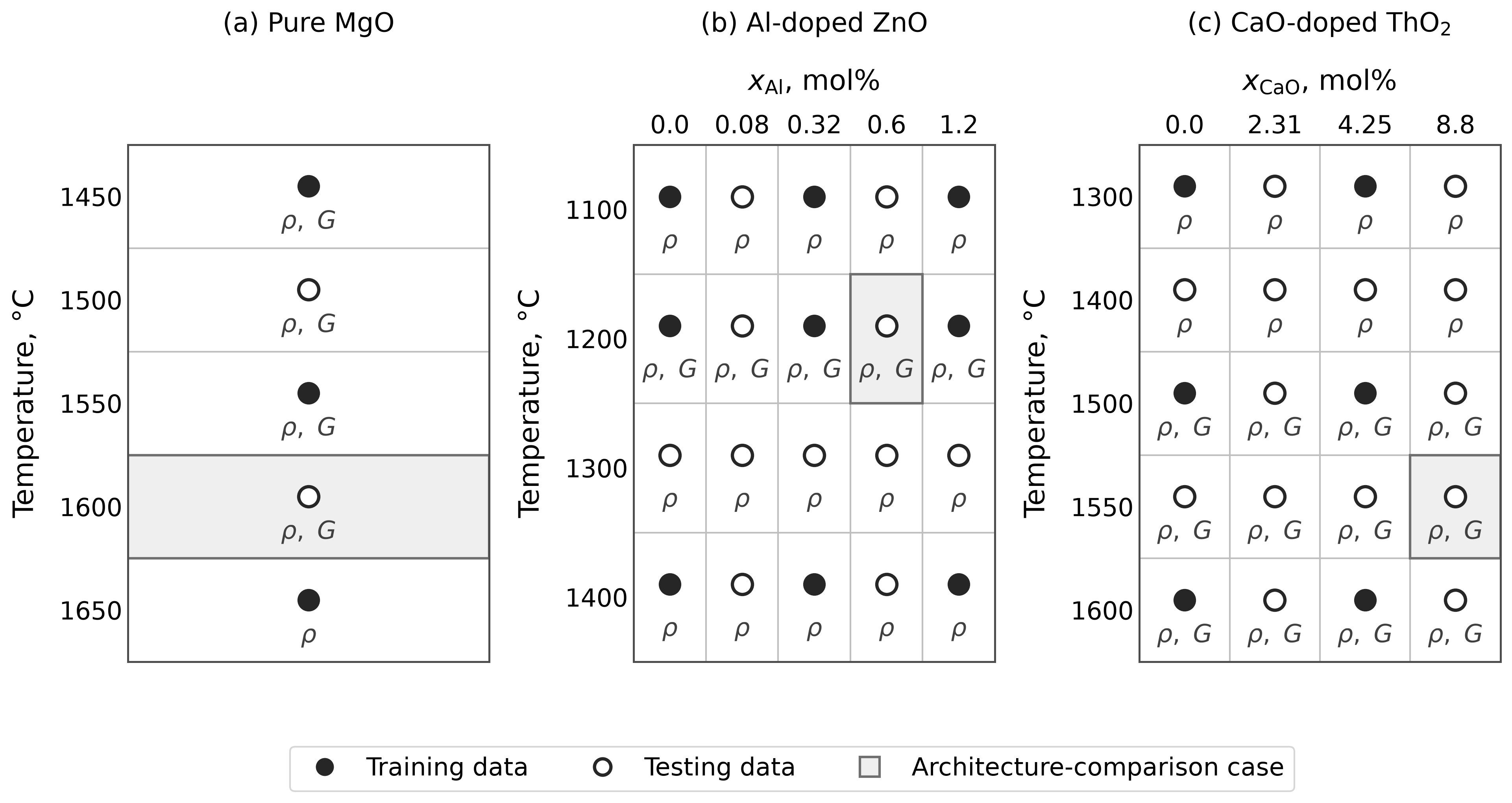}
    \caption{\textbf{Experimental design and data partitioning for the three material systems.} Filled and open circles denote training and testing conditions, respectively; $\rho$ and $G$ indicate the availability of density and grain-size measurements. Shaded cells identify the testing conditions used for the architecture rollout comparison: $1600\,^{\circ}\mathrm{C}$ for MgO, $0.60$ mol\% Al at $1200\,^{\circ}\mathrm{C}$ for Al-doped ZnO, and $8.8$ mol\% CaO at $1550\,^{\circ}\mathrm{C}$ for CaO-doped ThO$_2$. These conditions were selected independently of model performance as the testing conditions containing time-resolved measurements of both outputs that maximized the normalized distance from the corresponding training design. For MgO, the higher-temperature condition was selected to resolve the equal-distance case and evaluate rapid coupled kinetics beyond the highest grain-size-supervised training temperature.}
    \label{fig:experimental-design}
\end{figure}

Complete experimental cases, rather than individual time points from the same trajectory, were assigned to either training or testing. The resulting partition is shown in \cref{fig:experimental-design}; filled symbols denote training cases, open symbols denote testing cases, and the labels beneath the symbols indicate whether density alone or both density and grain size were available. This condition-wise design prevents later observations from a testing trajectory from leaking into training and directly evaluates generalization across processing conditions.

For MgO, the $1450$, $1550$, and $1650~^{\circ}\mathrm{C}$ conditions were used for training, while the interleaved $1500$ and $1600~^{\circ}\mathrm{C}$ conditions were reserved for testing. This partition tests temperature interpolation while requiring the model to learn from a training set with incomplete grain-size supervision. In particular, the grain-size prediction at $1600~^{\circ}\mathrm{C}$ lies above the highest temperature with direct grain-size supervision, even though density measurements at $1650~^{\circ}\mathrm{C}$ are included in training.

For Al-doped ZnO, the $0$, $0.32$, and $1.20$~mol\% Al conditions at $1100$, $1200$, and $1400~^{\circ}\mathrm{C}$ were used for training. The intermediate compositions of $0.08$ and $0.60$~mol\% at the same temperatures were withheld to test composition interpolation. All five compositions at $1300~^{\circ}\mathrm{C}$ were withheld to test temperature interpolation. This partition evaluates whether the coupled model can use limited grain-size histories at one temperature together with sparse density-only constraints at the other temperatures to predict composition-dependent densification and grain growth.

For CaO-doped $\mathrm{ThO_2}$, the undoped and $4.25$~mol\% CaO conditions at $1300$, $1500$, and $1600~^{\circ}\mathrm{C}$ were used for training. The $2.31$ and $8.80$~mol\% conditions at these temperatures were withheld to test composition interpolation and extrapolation, respectively. All compositions at $1400$ and $1550~^{\circ}\mathrm{C}$ were withheld to test temperature interpolation. This is the most demanding partition because most temperature--composition combinations are excluded from training while the model must reproduce both coupled observables where they are available.

%%%%%%%%%%%%%%%%%%%%%%%%%%%%%%%%%%%%%%%%%%%%%%%%%%%%%%%%%%%%%%%%%%%%%%%%%%%%%%%%%%%%%%%%%%%%%%%%%%%%%%%%%%%%%%%%%%%%%%%%%%%%%%%%%%%%%%%%
%%%%%%%%%%%%%%%%%%%%%%%%%%%%%%%%%%%%%%%%%%%%%%%%%%%%%%%%%%%%%%%%%%%%%%%%%%%%%%%%%%%%%%%%%%%%%%%%%%%%%%%%%%%%%%%%%%%%%%%%%%%%%%%%%%%%%%%%
\subsection{Data-driven baseline models}

Sinter-PiNDiff was compared with two data-driven architectures representing pointwise regression and residual state evolution. The multilayer perceptron (MLP)~\cite{Rumelhart1986} directly maps elapsed time and processing conditions, $[t,T,\chi_{\mathrm{d}},\chi_{\mathrm{b}}]$, to the corresponding density and grain size, $[\rho(t),G(t)]$. The residual network (ResNet) adopts the residual-learning formulation of He et al.~\cite{He2016ResNet}. At numerical step $n$, it maps the current density, processing conditions, and absolute time, $[\rho_n,T,\chi_{\mathrm{d}},\chi_{\mathrm{b}},t_n]$, to the nonnegative rates $[\dot{\rho}_n,\dot{G}_n]$. The full state is then advanced according to
\begin{equation}
    \bm{v}_{n+1}
    =
    \bm{v}_n
    +
    \Delta t_n
    \begin{bmatrix}
        \dot{\rho}_n \\
        \dot{G}_n
    \end{bmatrix}.
    \label{eq:resnet_update}
\end{equation}
The ResNet provides the closest data-driven analogue of Sinter-PiNDiff because both models evolve density and grain size autoregressively. However, the ResNet learns the state derivatives directly, whereas Sinter-PiNDiff learns effective kinetic coefficients that are evaluated within the retained sintering equations.

The learned mappings and trainable parameter counts of the three architectures are summarized in \cref{tab:model_formulations}. The table distinguishes the pointwise prediction used by the MLP, the learned state evolution used by the ResNet, and the learned kinetic closures embedded within Sinter-PiNDiff.

\begin{table}[ht]
\centering
\small
\caption{Learned mappings and trainable parameter counts for the architectures included in the model comparison. Parameter counts correspond to one independently trained model and are identical across the three material systems.}
\label{tab:model_formulations}
\setlength{\tabcolsep}{5pt}
\renewcommand{\arraystretch}{1.12}
\begin{tabular}{@{}p{0.20\linewidth}p{\dimexpr0.65\linewidth-4\tabcolsep\relax}>{\raggedleft\arraybackslash}p{0.15\linewidth}@{}}
\toprule
\textbf{Model} & \textbf{Learned mapping} & \textbf{Parameters} \\
\midrule
MLP &
$[t,T,\chi_{\mathrm{d}},\chi_{\mathrm{b}}]
\mapsto
[\rho(t),G(t)]$ &
55,014 \\
ResNet &
$[\rho_n,T,\chi_{\mathrm{d}},\chi_{\mathrm{b}},t_n]
\mapsto
[\dot{\rho}_n,\dot{G}_n]$ &
55,174 \\
Sinter-PiNDiff &
$[T,\rho_n,\chi_{\mathrm{d}},\chi_{\mathrm{b}}]
\mapsto
[D_{\mathrm{nn},n},k_{\mathrm{nn},n}]$, followed by the retained sintering equations &
5,538 \\
\bottomrule
\end{tabular}
\end{table}

The MLP and ResNet contain approximately ten times as many trainable parameters as Sinter-PiNDiff. The comparison is therefore not parameter matched. Nevertheless, the smaller parameter count of Sinter-PiNDiff shows that its improved predictive performance cannot be explained by a larger parameter budget. Instead, the comparison evaluates whether embedding the sintering equations provides a more effective inductive structure for learning coupled density and grain-size evolution from the available data. Complete layer structures, activation functions, output transformations, and scaling factors are provided in Supplementary Tables S1--S4.

All frameworks used the same condition-wise training and testing partitions shown in \cref{fig:experimental-design} and the same available-observation masks. Each model was optimized using Adam with cosine learning-rate decay, full-batch training for 2000 epochs, an $L_2$ regularization coefficient of $10^{-7}$, final-epoch model selection, and 64-bit numerical precision. The initial learning rate was $10^{-3}$ for the MLP and ResNet and $10^{-2}$ for Sinter-PiNDiff. Each framework was trained using five independent parameter initializations. Preprocessing parameters were determined from the training data or prescribed before evaluation, and no measurements from withheld temperature--composition conditions were used for model fitting or hyperparameter selection.

%%%%%%%%%%%%%%%%%%%%%%%%%%%%%%%%%%%%%%%%%%%%%%%%%%%%%%%%%%%%%%%%%%%%%%%%%%%%%%%%%%%%%%%%%%%%%%%%%%%%%%%%%%%%%%%%%%%%%%%%%%%%%%%%%%%%%%%%
%%%%%%%%%%%%%%%%%%%%%%%%%%%%%%%%%%%%%%%%%%%%%%%%%%%%%%%%%%%%%%%%%%%%%%%%%%%%%%%%%%%%%%%%%%%%%%%%%%%%%%%%%%%%%%%%%%%%%%%%%%%%%%%%%%%%%%%%

\subsection{Density-feedback ablation}
\label{sec:density-feedback-ablation-method}

An ablation study was performed to isolate the contribution of the evolving density to the learned kinetic closures. In the full Sinter-PiNDiff model, the effective densification and grain-growth coefficients are evaluated as
\begin{align}
    D_{\mathrm{nn},n}
    &=
    \mathcal{N}_{\rho}
    \left(
    T,\rho_n,\chi_{\mathrm{d}},\chi_{\mathrm{b}}
    \right),\\
    k_{\mathrm{nn},n}
    &=
    \mathcal{N}_{G}
    \left(
    T,\rho_n,\chi_{\mathrm{d}},\chi_{\mathrm{b}}
    \right).
\end{align}
For the ablated model, $\rho_n$ was removed from both neural-network inputs,
\begin{align}
    D_{\mathrm{nn},n}^{(-\rho)}
    &=
    \mathcal{N}_{\rho}^{(-\rho)}
    \left(
    T,\chi_{\mathrm{d}},\chi_{\mathrm{b}}
    \right),\\
    k_{\mathrm{nn},n}^{(-\rho)}
    &=
    \mathcal{N}_{G}^{(-\rho)}
    \left(
    T,\chi_{\mathrm{d}},\chi_{\mathrm{b}}
    \right).
\end{align}
All other components were unchanged, including the retained rate equations, hidden-layer architecture, density-saturation factor, numerical time steps, training--testing partitions, loss function, optimization schedule, and five-member ensemble protocol. Thus, the ablation isolates the learned dependence of the kinetic coefficients on the evolving density; it does not remove the prescribed saturation factor $\psi(\rho)$ from the densification equation.

For material system $s$ and error metric $M$, the ablation effect was quantified using
\begin{equation}
    R_{M}^{(s)}
    =
    \frac{
    \overline{E}_{M,\mathrm{without}\ \rho}^{(s)}
    }{
    \overline{E}_{M,\mathrm{full}}^{(s)}
    },
    \label{eq:ablation-error-ratio}
\end{equation}
where $\overline{E}$ denotes the mean error across the five independently initialized models. A ratio greater than one indicates that the full model has lower error, whereas a ratio below one indicates that the ablated model has lower error. Ratios were calculated for density NRMSE, grain-size NRMSE, final-density MAE, and final-grain-size MAPE using the same testing observations and metric definitions as the main model comparison.

%%%%%%%%%%%%%%%%%%%%%%%%%%%%%%%%%%%%%%%%%%%%%%%%%%%%%%%%%%%%%%%%%%%%%%%%%%%%%%%%%%%%%%%%%%%%%%%%%%%%%%%%%%%%%%%%%%%%%%%%%%%%%%%%%%%%%%%%
%%%%%%%%%%%%%%%%%%%%%%%%%%%%%%%%%%%%%%%%%%%%%%%%%%%%%%%%%%%%%%%%%%%%%%%%%%%%%%%%%%%%%%%%%%%%%%%%%%%%%%%%%%%%%%%%%%%%%%%%%%%%%%%%%%%%%%%%

\subsection{Evaluation protocol and error metrics}

Predictive accuracy was evaluated separately for the time-dependent trajectories and final states using normalized root-mean-square error (NRMSE), mean absolute error (MAE), and mean absolute percentage error (MAPE). For the following definitions, $\rho$ denotes relative density expressed as a fraction. For each material system, let $q_{c,i}$ and $\widehat{q}^{(m)}_{c,i}$ denote the measured and predicted values of observable $q\in\{\rho,G\}$ at observation $i$ in test condition $c$ for independently trained model $m$. Let $N_{q,c}$ denote the number of available observations of $q$ in condition $c$, and let $C_q$ denote the number of test conditions containing measurements of that observable.

Trajectory error was calculated by pooling the squared residuals across all available test observations:
\begin{equation}
    \mathrm{NRMSE}^{(m)}_{q}
    =
    \frac{100}{
        q_{\max,\mathrm{test}}-q_{\min,\mathrm{test}}
    }
    \sqrt{
        \frac{
            \displaystyle\sum_{c=1}^{C_q}
            \sum_{i=1}^{N_{q,c}}
            \left(
            \widehat{q}^{(m)}_{c,i}-q_{c,i}
            \right)^{2}
        }{
            \displaystyle\sum_{c=1}^{C_q}N_{q,c}
        }
    },
    \qquad q\in\{\rho,G\},
    \label{eq:nrmse}
\end{equation}
where $q_{\max,\mathrm{test}}$ and $q_{\min,\mathrm{test}}$ are the maximum and minimum measured test values of the corresponding observable and material system. NRMSE assesses agreement over the sampled histories while emphasizing larger discrepancies through its squared-error formulation~\cite{Chai2014}. Normalization by the observed test-output range removes the physical units and expresses the error relative to the variation within each test dataset~\cite{Botchkarev2019}. The same normalization is applied to all architectures within a material system. Because each observation receives equal weight, conditions containing more measurements contribute more strongly to the trajectory metric. Conditions containing only an endpoint measurement contribute that observation but do not independently establish accuracy throughout the intervening trajectory.

Final-state accuracy was evaluated separately because a trajectory-averaged error can obscure discrepancies at the endpoint. The final state is defined as the last experimentally reported value of the corresponding observable in each test condition. Final-density MAE was calculated as
\begin{equation}
    \mathrm{MAE}^{(m)}_{\rho,\mathrm{f}}
    =
    \frac{100}{C_{\rho}}
    \sum_{c=1}^{C_{\rho}}
    \left|
    \widehat{\rho}^{(m)}_{c,\mathrm{f}}
    -
    \rho_{c,\mathrm{f}}
    \right|.
    \label{eq:mae_density_final}
\end{equation}
Multiplication by $100$ converts the absolute difference in relative density to percentage points. MAE provides a directly interpretable measure of the discrepancy in achieved densification and places less emphasis on isolated large errors than a squared-error metric~\cite{Chai2014}. Because relative density already has a common physical scale, no additional normalization is applied.

Final grain-size accuracy was quantified using
\begin{equation}
    \mathrm{MAPE}^{(m)}_{G,\mathrm{f}}
    =
    \frac{100}{C_G}
    \sum_{c=1}^{C_G}
    \left|
    \frac{
        \widehat{G}^{(m)}_{c,\mathrm{f}}
        -
        G_{c,\mathrm{f}}
    }{
        G_{c,\mathrm{f}}
    }
    \right|,
    \label{eq:mape_grain_final}
\end{equation}
and is reported as a percentage of the measured final grain size. MAPE expresses endpoint errors in proportional terms and is useful when the magnitude of the target varies across conditions and material systems~\cite{deMyttenaere2016}. Unlike NRMSE, which uses a common output range within each material-specific test dataset, MAPE normalizes each endpoint error by its corresponding measured value. Smaller reference values therefore receive greater weight for a fixed absolute discrepancy, and zero reference values are inadmissible~\cite{deMyttenaere2016}. Each test condition contributes one endpoint per available observable, so the final-state metrics weight the eligible conditions equally.

The three metrics provide complementary assessments: NRMSE summarizes discrepancies across the sampled histories, final-density MAE quantifies the absolute error in achieved relative density, and final-grain-size MAPE quantifies the proportional error in the resulting grain size. Each metric was calculated separately for every independently trained model and then reported as the mean $\pm$ one sample standard deviation across the five initializations. The reported standard deviations therefore characterize variability in predictive accuracy across independent parameter initializations rather than experimental uncertainty.

Empirical coverage of the ensemble bands was evaluated over the experimental test observations. For each material system and observable, the coverage was calculated as

\begin{equation}
\mathrm{Cov}_{k,q}
=
\frac{1}{N_{q,\mathrm{test}}}
\sum_{i=1}^{N_{q,\mathrm{test}}}
\mathbf{1}
\left(
\left|q_i^{\mathrm{exp}}-\overline{q}_i\right|
\leq k\sigma_{q,i}
\right),
\qquad k\in\{2,3\},
\label{eq:empirical_coverage}
\end{equation}

where $N_{q,\mathrm{test}}$ is the number of test observations for output $q$, $\mathbf{1}(\cdot)$ is the indicator function, and $\overline{q}_i$ and $\sigma_{q,i}$ are the ensemble mean and standard deviation at the corresponding experimental condition and time. Coverage was calculated separately for density and grain size using the same test observations included in the trajectory-error metrics. Prescribed initial conditions were not counted as prediction targets. Because the ensemble spread does not include a separately modeled contribution from experimental noise, these values quantify the empirical coverage of the ensemble-disagreement bands rather than calibrated confidence or prediction intervals.

%%%%%%%%%%%%%%%%%%%%%%%%%%%%%%%%%%%%%%%%%%%%%%%%%%%%%%%%%%%%%%%%%%%%%%%%%%%%%%%%%%%%%%%%%%%%%%%%%%%%%%%%%%%%%%%%%%%%%%%%%%%%%%%%%%%%%%%%
%%%%%%%%%%%%%%%%%%%%%%%%%%%%%%%%%%%%%%%%%%%%%%%%%%%%%%%%%%%%%%%%%%%%%%%%%%%%%%%%%%%%%%%%%%%%%%%%%%%%%%%%%%%%%%%%%%%%%%%%%%%%%%%%%%%%%%%%%%%%%%%%%%%%%%%%%%%%%%%%%%%%%%%%%%%%%%%%%%%%%%%%%%%%%%%%%%%%%%%%%%%%%%%%%%%%%%%%%%%%%%%%%%%%%%%%%%%%%%%%%%%%%%%%%%%%%%%%%%%%%%%%%%%%%%%%

\section{Results}

The evaluation addresses four complementary questions: whether the same physics-integrated formulation can be retrained for distinct material systems, whether its ensemble spread reflects differences in experimental coverage, whether the retained sintering equations improve predictions relative to data-driven alternatives, and whether supplying the evolving density to the learned kinetic closures improves predictive accuracy. The common workflow and recurrent architecture in \cref{fig:sinter-pindiff-overview} and \cref{fig:sinter-pindiff-architecture} were applied separately to each material system using the training and testing dataset in \cref{fig:experimental-design}. Here, retrainability means that the same governing equations, neural-network architecture, and optimization procedure are fitted anew using data from each material system. \Cref{fig:mgo_temporal_predictions,fig:al_zno_1200c,fig:al_zno_120min,fig:tho2_density,fig:tho2_grain} assess the resulting predictions across the available processing conditions, whereas \cref{tab:pindiff_baseline_comparison} and \cref{fig:mgo_model_comparison,fig:zno_model_comparison,fig:tho2_model_comparison} evaluate the contribution of the physics-integrated formulation against the two baselines. \Cref{fig:density-feedback-ablation} evaluates the contribution of density-dependent kinetic feedback.

%%%%%%%%%%%%%%%%%%%%%%%%%%%%%%%%%%%%%%%%%%%%%%%%%%%%%%%%%%%%%%%%%%%%%%%%%%%%%%%%%%%%%%%%%%%%%%%%%%%%%%%%%%%%%%%%%%%%%%%%%%%%%%%%%%%%%%%%

\subsection{Applicability across material systems}
\label{sec:material-applicability}

%%%%%%%%%%%%%%%%%%%%%%%%%%%%%%%%%%%%%%%%%%%%%%%%%%%%%%%%%%%%%%%%%%%%%%%%%%%%%%%%%%%%%%%%%%%%%%%%%%%%%%%%%%%%%%%%%%%%%%%%%%%%%%%%%%%%%%%%

\subsubsection{MgO}

\begin{figure}[ht]
    \centering
    \includegraphics
    [width=\textwidth]{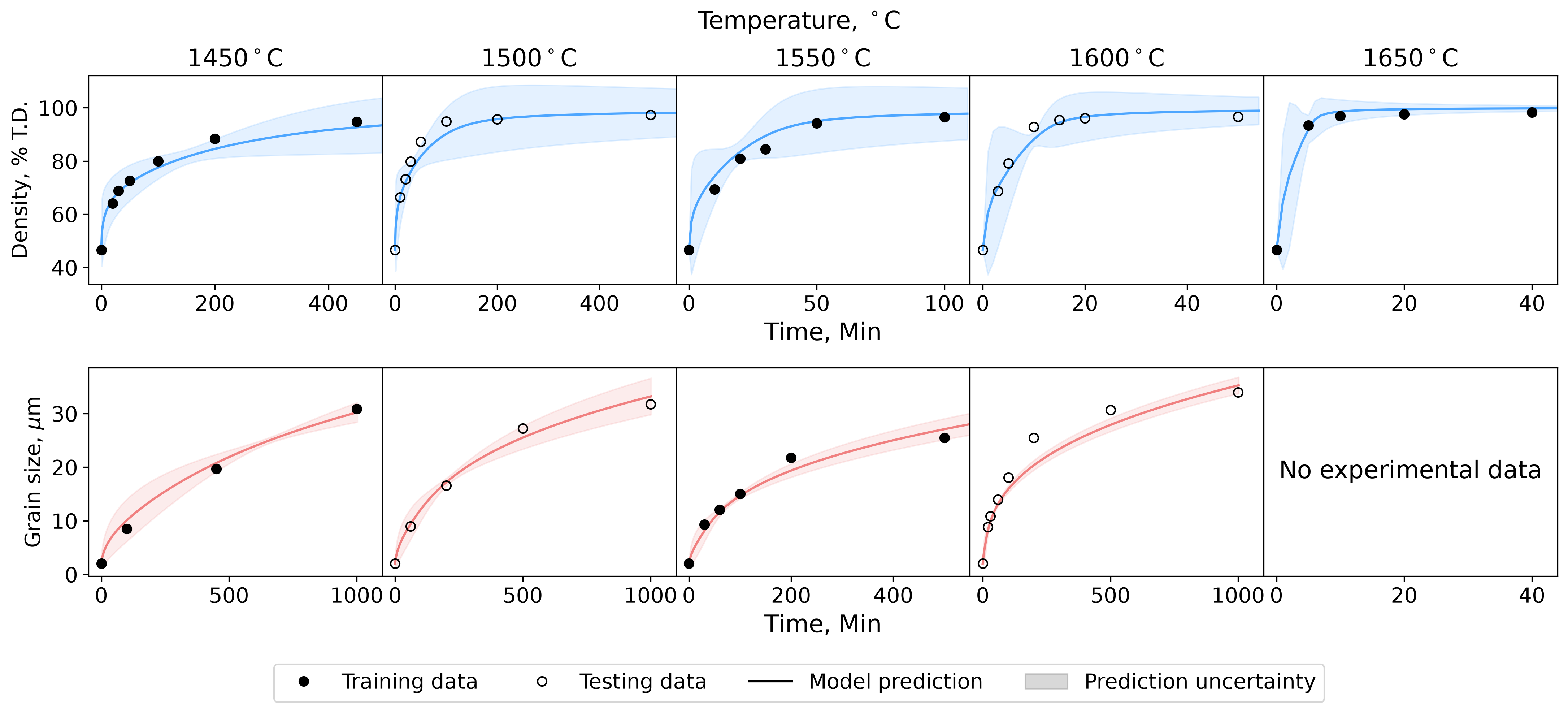}
    \caption{\textbf{Temporal predictions for MgO across the investigated sintering temperatures.} The upper and lower rows show density and grain-size evolution, respectively. Filled and open circles denote training and testing experimental data. Solid curves show the five-model ensemble mean, and shaded bands indicate $\pm3$ ensemble standard deviations. No experimental grain-size measurements were available at $1650\,^{\circ}\mathrm{C}$.}
    \label{fig:mgo_temporal_predictions}
\end{figure}

\Cref{fig:mgo_temporal_predictions} shows that Sinter-PiNDiff reproduces the rapid densification and continued grain growth in MgO. Relative density rises rapidly at the start of the hold and subsequently approaches a plateau, while grain size continues to increase over a substantially longer interval. The predicted density histories also capture the acceleration of densification with increasing temperature. At the unseen temperatures of $1500$ and $1600\,^{\circ}\mathrm{C}$, the ensemble means closely follow the experimental data. Agreement is not uniform at every observation: early-to-intermediate density values at $1600\,^{\circ}\mathrm{C}$ are underpredicted, and some intermediate grain-size measurements lie above the mean. Nevertheless, the framework recovers the distinct shapes and characteristic time scales of both observables across the temperature series.

The observation pattern is important when interpreting uncertainty. For density, both test temperatures lie between training conditions. For grain size, $1500\,^{\circ}\mathrm{C}$ is bracketed by measured training histories at $1450$ and $1550\,^{\circ}\mathrm{C}$, whereas $1600\,^{\circ}\mathrm{C}$ lies above the highest temperature with grain-size supervision. Although density measurements at $1650\,^{\circ}\mathrm{C}$ enter the coupled training problem, they do not provide direct grain-size labels. The broader late-time grain-size band at $1600\,^{\circ}\mathrm{C}$ relative to $1500\,^{\circ}\mathrm{C}$ is therefore consistent with weaker constraints on the growth kinetics beyond the grain-size training range. No grain-size validation is possible at $1650\,^{\circ}\mathrm{C}$ because measurements were unavailable. This case demonstrates that interpolation and extrapolation must be interpreted with respect to the available observations of each output, even when both outputs share a coupled model.

%%%%%%%%%%%%%%%%%%%%%%%%%%%%%%%%%%%%%%%%%%%%%%%%%%%%%%%%%%%%%%%%%%%%%%%%%%%%%%%%%%%%%%%%%%%%%%%%%%%%%%%%%%%%%%%%%%%%%%%%%%%%%%%%%%%%%%%%

\FloatBarrier
\subsubsection{Al-doped ZnO}

\begin{figure}[ht]
    \centering
    \includegraphics[width=\textwidth]{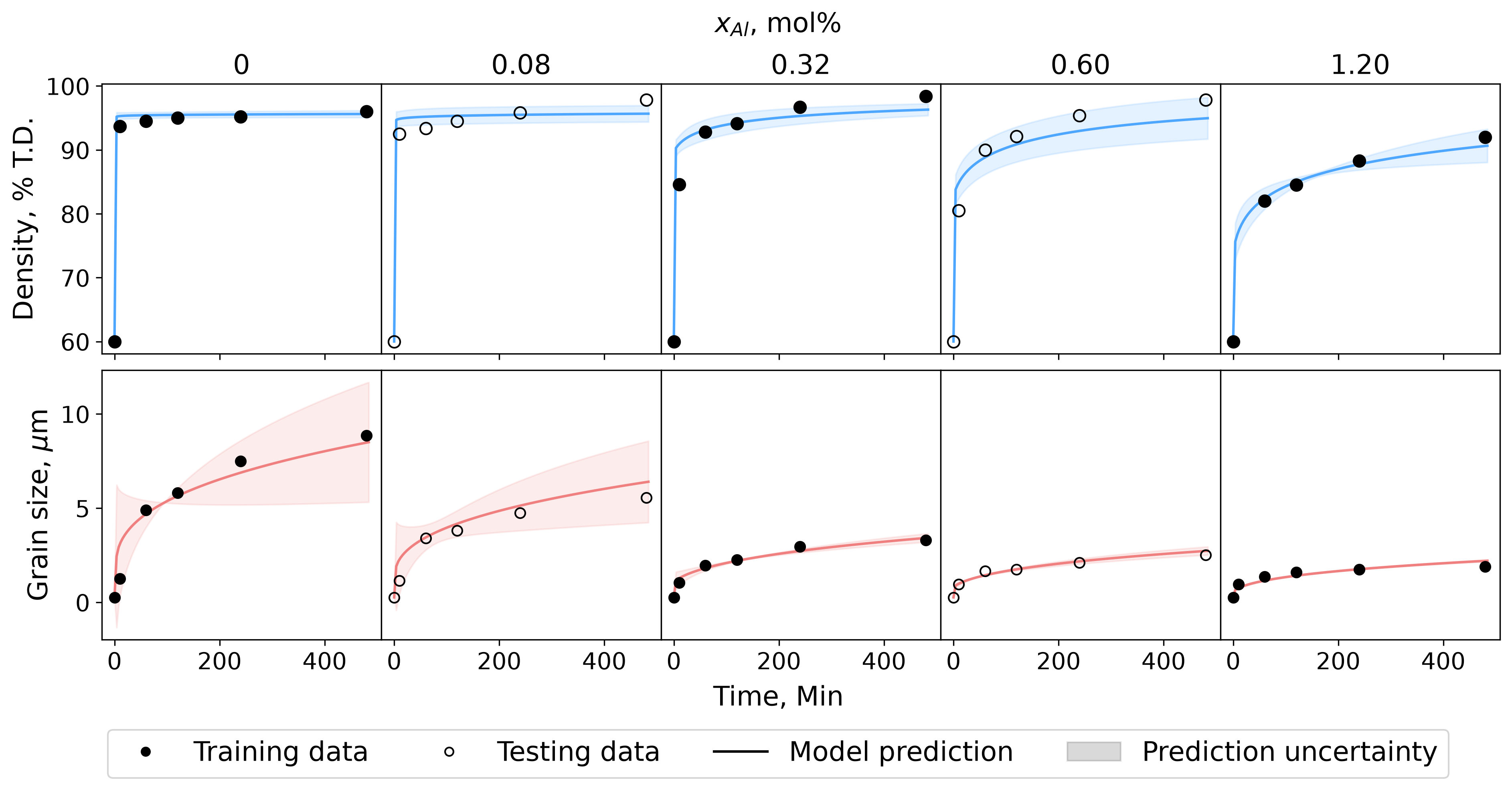}
    \caption{\textbf{Composition-dependent temporal predictions for Al-doped ZnO at $1200\,^{\circ}\mathrm{C}$.} Density and grain-size evolution are shown in the upper and lower rows, respectively, for Al contents from $0$ to $1.20$ mol\%. The intermediate $0.08$ and $0.60$ mol\% compositions were withheld for testing. Filled and open circles denote training and testing measurements. Solid curves show the five-model ensemble mean, and shaded bands indicate $\pm3$ ensemble standard deviations.}
    \label{fig:al_zno_1200c}
\end{figure}

\Cref{fig:al_zno_1200c} tests whether the same formulation can represent composition-dependent sintering at $1200\,^{\circ}\mathrm{C}$. The experimental data span rapid densification in undoped ZnO, slower densification at higher Al contents, and a reduction in grain growth with Al addition. Sinter-PiNDiff captures these contrasting responses through composition-dependent kinetic closures while retaining the rate equations used for MgO. The predicted grain sizes decrease from the undoped condition toward the more heavily doped compositions, and the density curves reproduce the slower approach to high density at $0.60$ and $1.20$~mol\% Al.

The withheld $0.08$ and $0.60$~mol\% Al cases provide two distinct tests of composition interpolation. At $0.08$~mol\%, the model recovers the high-density trajectory and the intermediate grain-growth response between undoped and more strongly doped ZnO. At $0.60$~mol\%, it captures the slower densification and much smaller grain sizes, although the late-time density remains somewhat underestimated. The grain-size band is narrow at $0.60$~mol\% and broader at $0.08$~mol\%, reflecting different levels of agreement among the learned closures within the same interpolation domain. The broad grain-size spread for undoped ZnO, despite its inclusion in training, further shows that a training label alone does not imply tightly identified kinetics. These variations support an interpretation of ensemble spread as sensitivity to the information supplied by the data, rather than as a binary distinction between training and testing cases.

\begin{figure}[ht]
    \centering
    \includegraphics[width=\textwidth]{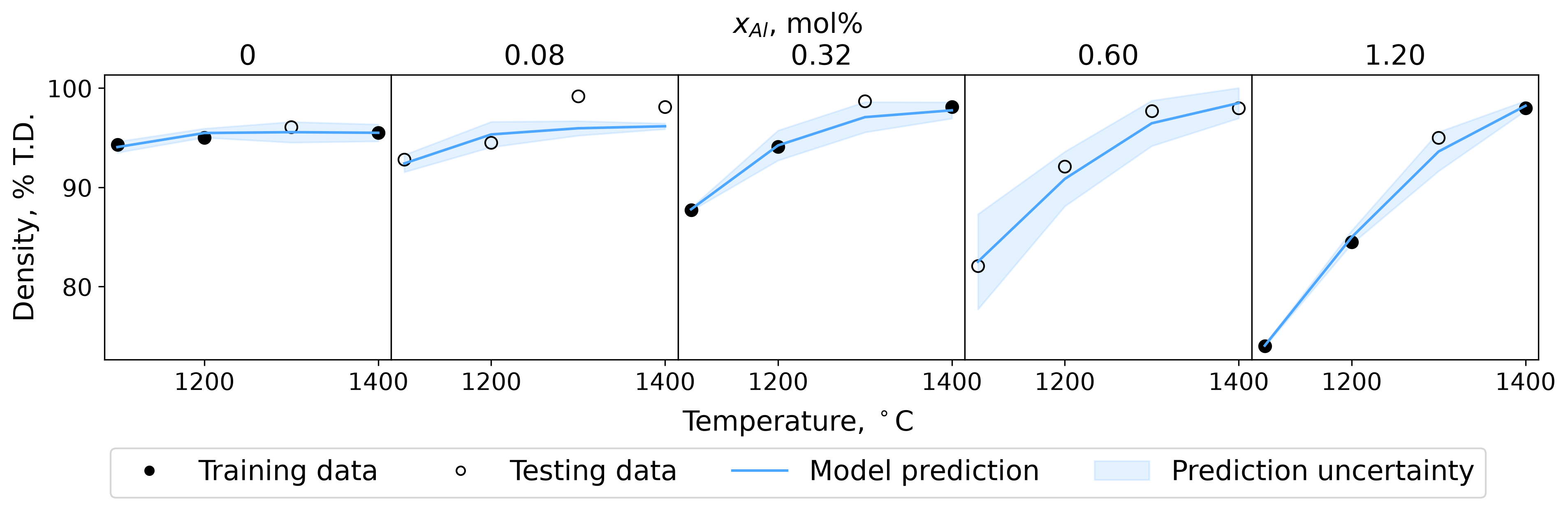}
    \caption{\textbf{Temperature-dependent density predictions for Al-doped ZnO after $120$ min of sintering.} Results are shown from $1100$ to $1400\,^{\circ}\mathrm{C}$ for Al contents from $0$ to $1.20$ mol\%. Filled and open circles denote training and testing measurements. Solid curves show the five-model ensemble mean, and shaded bands indicate $\pm3$ ensemble standard deviations.}
    \label{fig:al_zno_120min}
\end{figure}

\Cref{fig:al_zno_120min} extends the assessment to the relative density after $120$~min across temperature and composition. The undoped material shows relatively little change in final density over the investigated temperature interval, whereas the more heavily doped conditions exhibit a much stronger temperature dependence. The model reproduces this contrast: the predicted temperature response becomes steeper as Al content increases, and the high-temperature densities converge toward similarly high values. At the entirely withheld temperature of $1300\,^{\circ}\mathrm{C}$, the predictions broadly track the measured composition-dependent densities, with visible underprediction for some doped conditions. The $0.08$ and $0.60$~mol\% cases additionally test composition interpolation across the temperature interval. The wider band for $0.60$~mol\% near $1100\,^{\circ}\mathrm{C}$ illustrates that interpolation can remain uncertain when the response varies strongly between the available training compositions.

%%%%%%%%%%%%%%%%%%%%%%%%%%%%%%%%%%%%%%%%%%%%%%%%%%%%%%%%%%%%%%%%%%%%%%%%%%%%%%%%%%%%%%%%%%%%%%%%%%%%%%%%%%%%%%%%%%%%%%%%%%%%%%%%%%%%%%%%

\FloatBarrier
\subsubsection{\texorpdfstring{CaO-doped ThO$_2$}{CaO-doped ThO2}}

\begin{figure}[ht]
    \centering
    \includegraphics[width=0.9\textwidth]{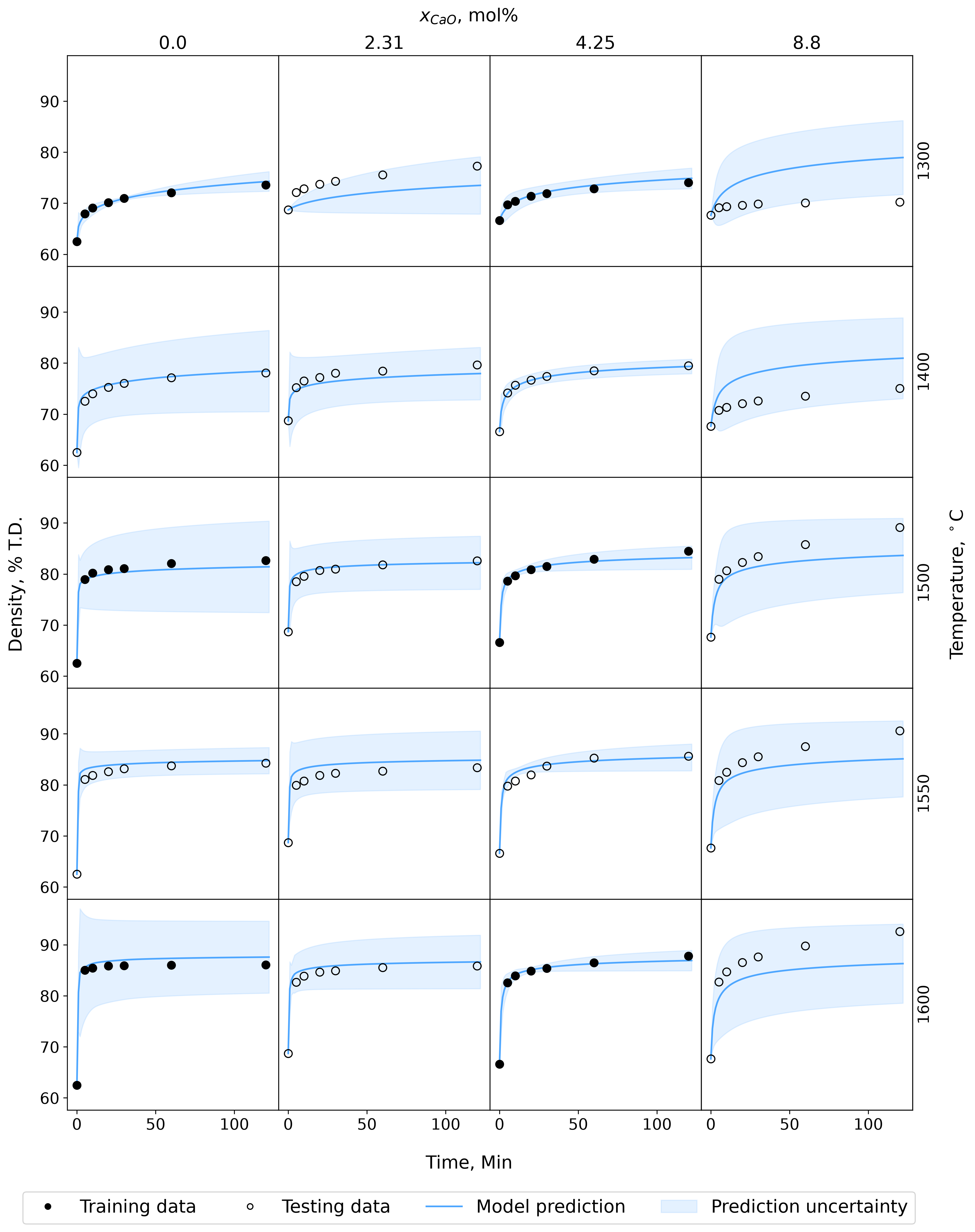}
    \caption{\textbf{Temporal density predictions for CaO-doped ThO$_2$.} Columns correspond to CaO contents of $0$, $2.31$, $4.25$, and $8.8$ mol\%, and rows correspond to temperatures from $1300$ to $1600\,^{\circ}\mathrm{C}$. Filled and open circles denote training and testing measurements. Solid curves show the five-model ensemble mean, and shaded bands indicate $\pm3$ ensemble standard deviations.}
    \label{fig:tho2_density}
\end{figure}

\Cref{fig:tho2_density} examines the most extensive withheld design: six of the twenty temperature--composition conditions were used for training, leaving fourteen for testing. For undoped ThO$_2$ and the $4.25$~mol\% CaO composition, the predicted density histories reproduce the rapid initial densification followed by a slower approach to a temperature-dependent plateau. The withheld $1400$ and $1550\,^{\circ}\mathrm{C}$ conditions test temperature interpolation, while the $2.31$~mol\% composition tests interpolation between the two training compositions. The ensemble means broadly follow the measured density evolution in these cases, including combinations for which both temperature and composition were withheld. This demonstrates that the common model can use a small set of complete trajectories to predict additional combinations of processing variables.

\begin{figure}[ht]
    \centering
    \includegraphics[width=\textwidth]{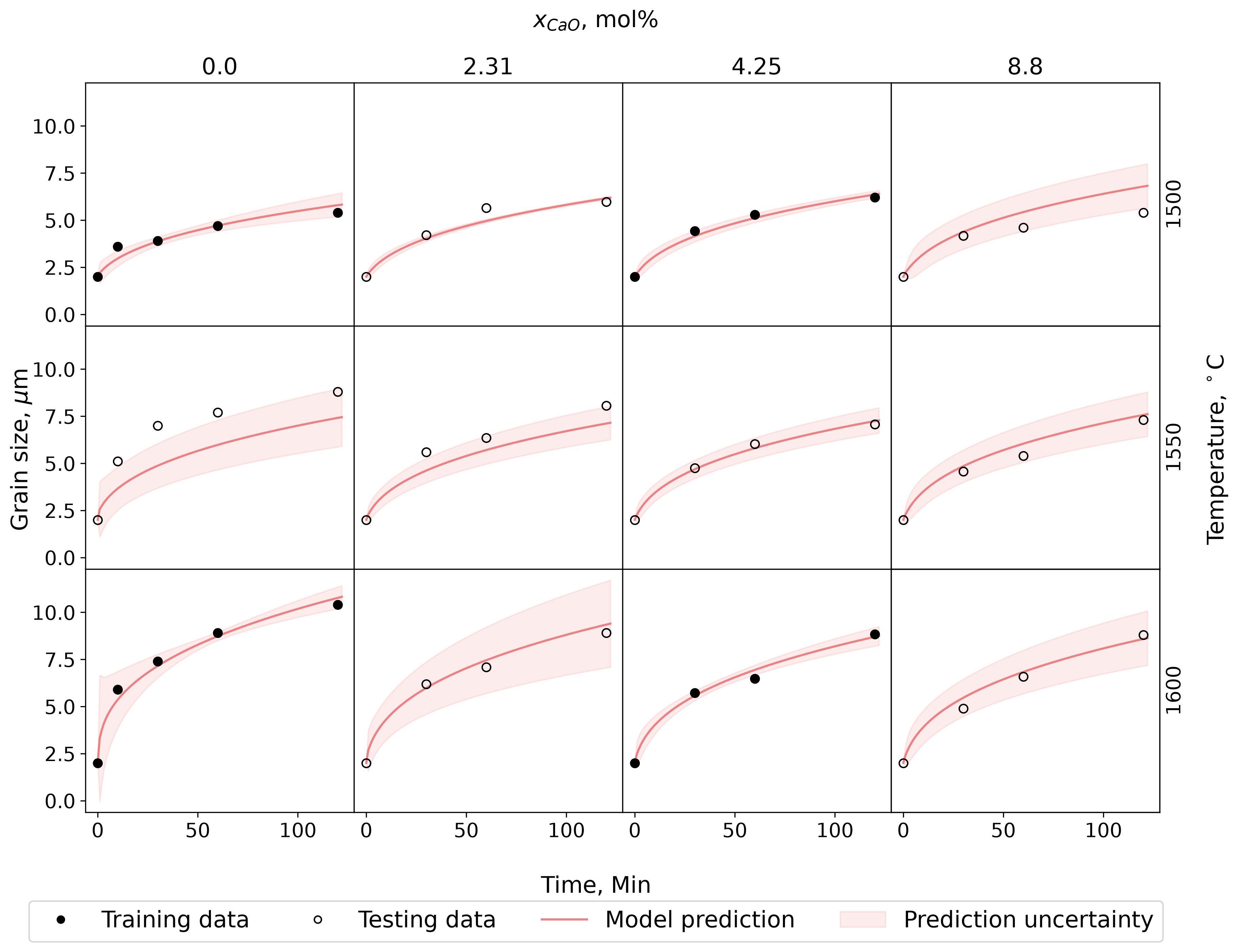}
    \caption{\textbf{Temporal grain-size predictions for CaO-doped ThO$_2$.} Columns correspond to CaO contents of $0$, $2.31$, $4.25$, and $8.8$ mol\%, and rows correspond to temperatures of $1500$, $1550$, and $1600\,^{\circ}\mathrm{C}$. Filled and open circles denote training and testing measurements. Solid curves show the five-model ensemble mean, and shaded bands indicate $\pm3$ ensemble standard deviations.}
    \label{fig:tho2_grain}
\end{figure}

\Cref{fig:tho2_grain} evaluates the grain-growth component over the temperatures for which grain-size measurements are available. The model reproduces the overall increase in grain size with hold time and the stronger coarsening at higher temperatures. Close agreement for the $4.25$~mol\% cases at the training temperatures is accompanied by useful predictions at the withheld $1550\,^{\circ}\mathrm{C}$ condition and at the intermediate $2.31$~mol\% composition. The latter is especially relevant because it tests a composition that contributes neither density nor grain-size observations to training. Some discrepancies remain, including underprediction of the undoped grain size at $1550\,^{\circ}\mathrm{C}$ and overprediction for $8.8$~mol\% CaO at $1500\,^{\circ}\mathrm{C}$. The extrapolated $8.8$~mol\% histories generally have broader bands than their $2.31$~mol\% counterparts, consistent with the pattern observed for density.

The $8.8$~mol\% CaO column provides a more demanding test because it lies outside the $0$--$4.25$~mol\% training-composition range. Its density bands are generally wider than those of the corresponding $2.31$~mol\% interpolation cases, particularly during the later portion of the hold. This increase in spread is consistent with the greater freedom of the neural closures outside the observed composition range. The mean predictions also reveal a systematic limitation: they tend to overestimate the low-temperature plateau and underestimate continued densification at the higher temperatures. At $1500$--$1600\,^{\circ}\mathrm{C}$, the predicted curves level off earlier than the experimental histories, so a late-time bias remains despite the enlarged bands. The extrapolation result therefore supports the usefulness of reporting model uncertainty together with the mean, while identifying a composition range where additional measurements or a richer kinetic representation would be valuable.

%%%%%%%%%%%%%%%%%%%%%%%%%%%%%%%%%%%%%%%%%%%%%%%%%%%%%%%%%%%%%%%%%%%%%%%%%%%%%%%%%%%%%%%%%%%%%%%%%%%%%%%%%%%%%%%%%%%%%%%%%%%%%%%%%%%%%%%%
%%%%%%%%%%%%%%%%%%%%%%%%%%%%%%%%%%%%%%%%%%%%%%%%%%%%%%%%%%%%%%%%%%%%%%%%%%%%%%%%%%%%%%%%%%%%%%%%%%%%%%%%%%%%%%%%%%%%%%%%%%%%%%%%%%%%%%%%
%%%%%%%%%%%%%%%%%%%%%%%%%%%%%%%%%%%%%%%%%%%%%%%%%%%%%%%%%%%%%%%%%%%%%%%%%%%%%%%%%%%%%%%%%%%%%%%%%%%%%%%%%%%%%%%%%%%%%%%%%%%%%%%%%%%%%%%%

\FloatBarrier
\subsection{Empirical coverage of ensemble bands}
\label{sec:ensemble_coverage}

The empirical coverage of the Sinter-PiNDiff ensemble bands was evaluated over all testing observations using \cref{eq:empirical_coverage}. As shown in \cref{tab:uncertainty_coverage}, coverage varied substantially across material systems and observables. For density, the $\pm2\sigma$ bands contained 76.9\%, 47.4\%, and 78.6\% of the MgO, Al-doped ZnO, and CaO-doped ThO$_2$ test observations, respectively. Expanding the bands to $\pm3\sigma$ increased the corresponding coverage to 84.6\%, 63.2\%, and 90.5\%.

\begin{table}[ht]
\centering
\small
\caption{Empirical coverage of the Sinter-PiNDiff ensemble bands over the testing observations. Entries report the number of covered observations divided by the total number of test observations, with the corresponding percentage in parentheses. Here, $\sigma$ denotes the standard deviation across five independently trained ensemble members.}
\label{tab:uncertainty_coverage}
\begin{tabular}{@{}llrcc@{}}
\toprule
\textbf{Material system} &
\textbf{Observable} &
\textbf{$N_{\mathrm{test}}$} &
\textbf{Within $\pm2\sigma$} &
\textbf{Within $\pm3\sigma$} \\
\midrule
MgO
& Density
& 13
& 10/13 (76.9\%)
& 11/13 (84.6\%) \\

& Grain size
& 11
& 2/11 (18.2\%)
& 6/11 (54.5\%) \\
\midrule

Al-doped ZnO
& Density
& 19
& 9/19 (47.4\%)
& 12/19 (63.2\%) \\

& Grain size
& 10
& 8/10 (80.0\%)
& 8/10 (80.0\%) \\
\midrule

CaO-doped ThO$_2$
& Density
& 84
& 66/84 (78.6\%)
& 76/84 (90.5\%) \\

& Grain size
& 25
& 13/25 (52.0\%)
& 16/25 (64.0\%) \\
\bottomrule
\end{tabular}
\end{table}

Grain-size coverage was generally lower and more variable. The $\pm2\sigma$ coverage ranged from 18.2\% for MgO to 80.0\% for Al-doped ZnO, while the corresponding $\pm3\sigma$ coverage ranged from 54.5\% to 80.0\%. The particularly low coverage for MgO grain size and Al-doped ZnO density shows that ensemble disagreement does not capture the full model--data discrepancy for these outputs. Several measurements remain outside even the $\pm3\sigma$ bands, indicating that the ensemble members can share a common prediction bias.

These results confirm that the reported bands should be interpreted as descriptive measures of disagreement among independently trained models rather than calibrated prediction intervals. The coverage values also depend on the number and distribution of available test observations; in particular, the grain-size results are based on only 10--25 test points per material system. Consequently, the values provide an empirical assessment of ensemble-band reliability for the present datasets but do not establish general coverage probabilities.

%%%%%%%%%%%%%%%%%%%%%%%%%%%%%%%%%%%%%%%%%%%%%%%%%%%%%%%%%%%%%%%%%%%%%%%%%%%%%%%%%%%%%%%%%%%%%%%%%%%%%%%%%%%%%%%%%%%%%%%%%%%%%%%%%%%%%%%%
%%%%%%%%%%%%%%%%%%%%%%%%%%%%%%%%%%%%%%%%%%%%%%%%%%%%%%%%%%%%%%%%%%%%%%%%%%%%%%%%%%%%%%%%%%%%%%%%%%%%%%%%%%%%%%%%%%%%%%%%%%%%%%%%%%%%%%%%
%%%%%%%%%%%%%%%%%%%%%%%%%%%%%%%%%%%%%%%%%%%%%%%%%%%%%%%%%%%%%%%%%%%%%%%%%%%%%%%%%%%%%%%%%%%%%%%%%%%%%%%%%%%%%%%%%%%%%%%%%%%%%%%%%%%%%%%%

\FloatBarrier
\subsection{Benchmarking against data-driven baselines}
\label{sec:baseline-results}

\Cref{tab:pindiff_baseline_comparison} summarizes performance over all available test observations and endpoints. Sinter-PiNDiff achieves the lowest mean error for every reported metric in all three material systems. Its density NRMSE is $14.6\pm3.5$\% for MgO, $10.8\pm1.4$\% for Al-doped ZnO, and $14.4\pm3.9$\% for CaO-doped ThO$_2$. Grain-size NRMSE also decreases in all three systems, reaching $8.6\pm0.7$\%, $12.1\pm4.1$\%, and $19.3\pm5.2$\%, respectively. These comparisons use the same test observations and normalization within each material system.

\begin{table}[ht]
    \centering
    \caption{\textbf{Predictive performance of Sinter-PiNDiff and data-driven baseline models.} Values are reported as the mean $\pm$ one sample standard deviation across five independent model initializations evaluated on the testing dataset. Each normalized root-mean-square error (NRMSE) pools all available observations of the corresponding output and is expressed as a percentage of its test-output range. Conditions with endpoint-only measurements contribute only those observations. Final-density mean absolute error (MAE) is reported in percentage points (p.p.), and final-grain-size mean absolute percentage error (MAPE) is reported as a percentage. Lower values indicate better performance; the lowest mean for each material system and metric is shown in bold.}
    \label{tab:pindiff_baseline_comparison}
    \includegraphics[
        width=\textwidth,
        keepaspectratio
    ]{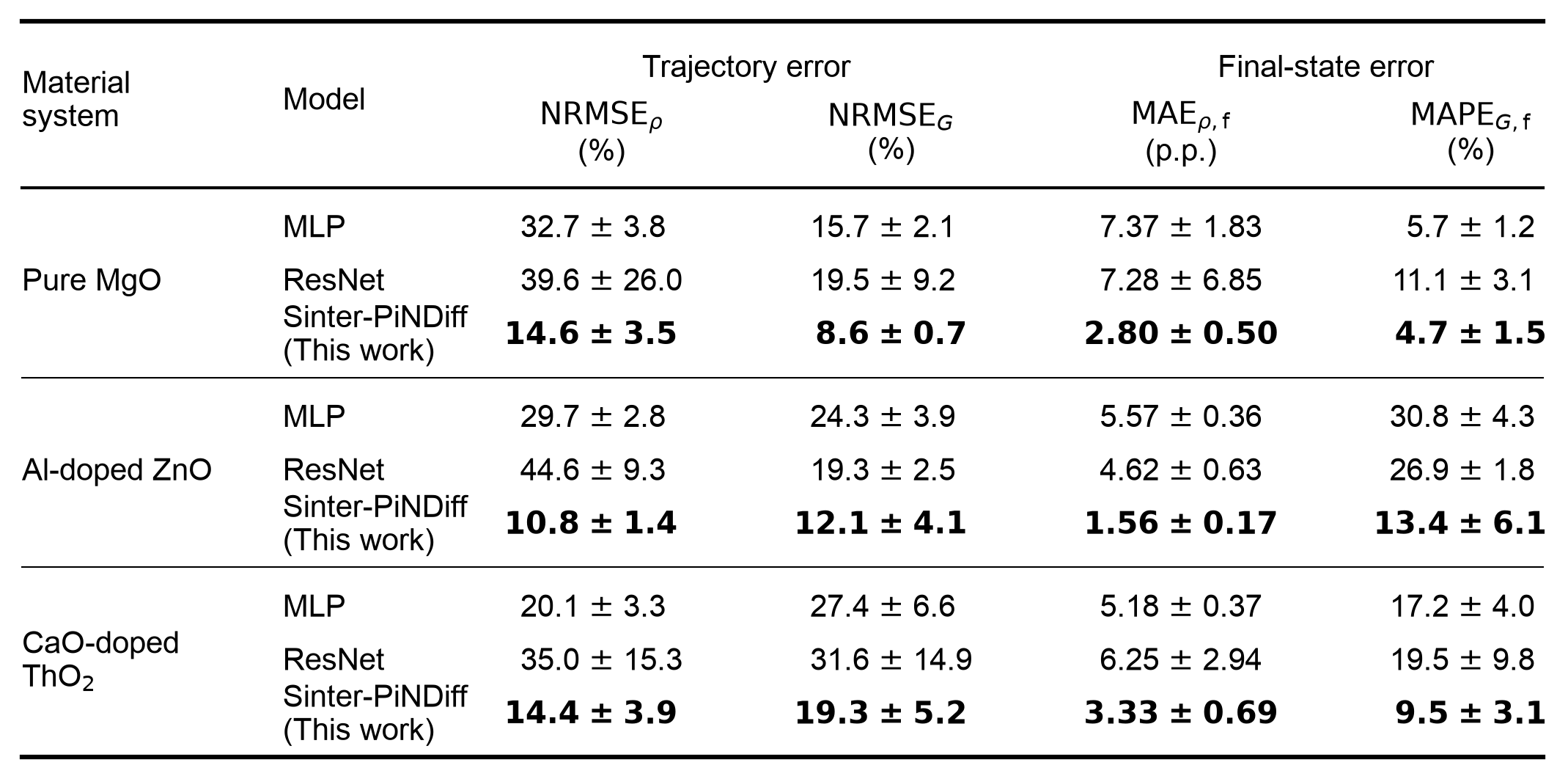}
\end{table}

The endpoint metrics show that the improvement extends to the final density and grain size. Final-density MAE decreases from $7.28$ to $2.80$ percentage points for MgO, from $4.62$ to $1.56$ percentage points for Al-doped ZnO, and from $5.18$ to $3.33$ percentage points for CaO-doped ThO$_2$, when compared with the better baseline mean for each system. Final-grain-size MAPE is also lowest for Sinter-PiNDiff, at $4.7$\%, $13.4$\%, and $9.5$\%, respectively. The MgO grain-size endpoint advantage over the MLP is modest ($4.7$\% versus $5.7$\%), illustrating why trajectory and endpoint errors should be considered together. The reported standard deviations quantify variation across five initializations; they do not establish statistical significance, and Sinter-PiNDiff does not have the smallest standard deviation in every comparison.

The numerical metrics are calculated for individual trained models and then averaged across initializations, whereas the following figures display the ensemble-mean trajectories and their spread. \Cref{fig:mgo_model_comparison,fig:zno_model_comparison,fig:tho2_model_comparison} therefore complement the aggregate table by showing where the architectures differ in temporal shape, bias, and agreement among ensemble members. Each figure uses the testing condition identified in \cref{fig:experimental-design}, rather than a condition selected on the basis of prediction accuracy.

\FloatBarrier
\subsubsection{MgO}

\begin{figure}[ht]
    \centering
    \includegraphics[width=0.9\textwidth]{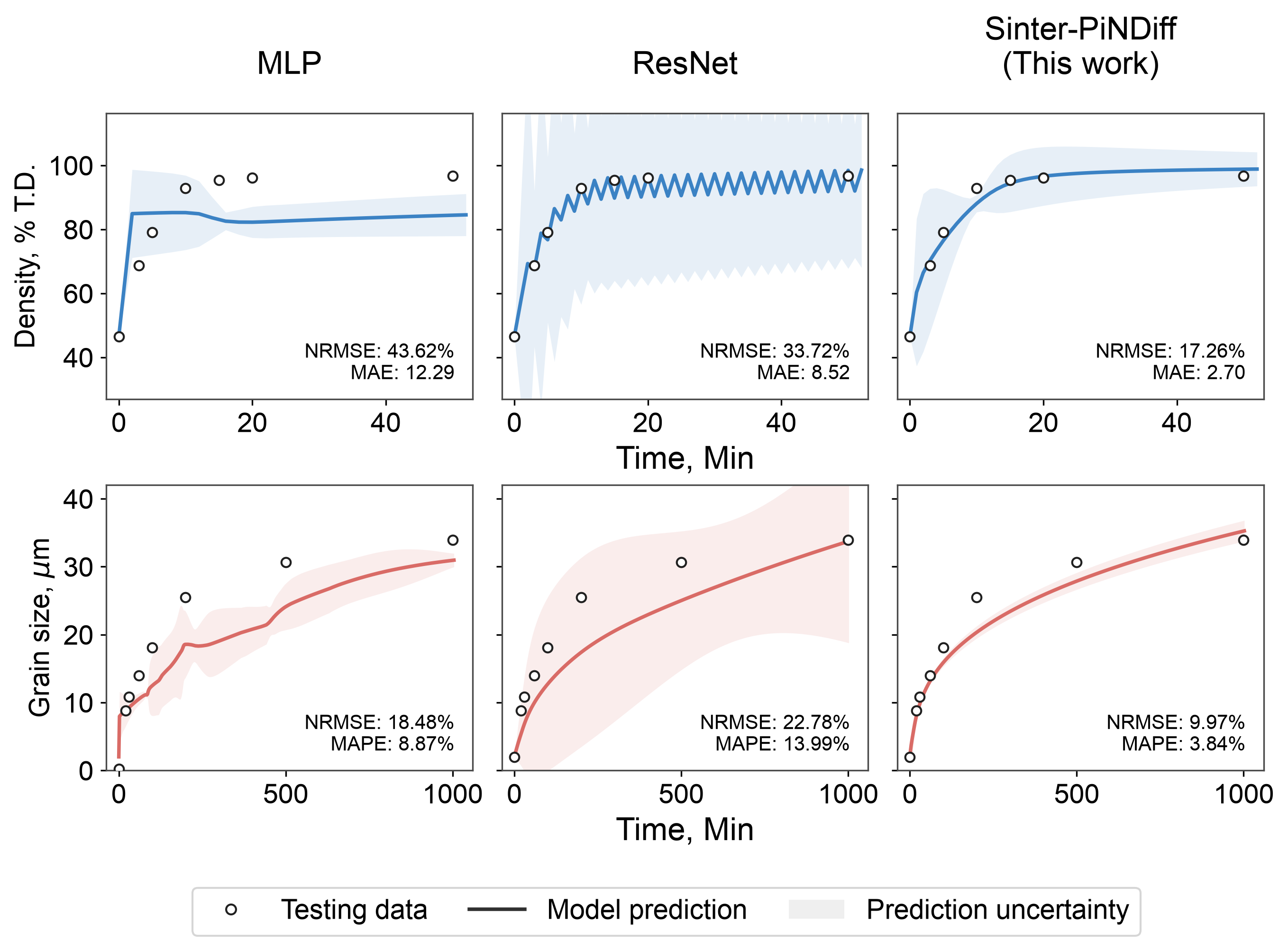}
    \caption{\textbf{Comparison of temporal predictions for MgO at $1600\,^{\circ}\mathrm{C}$.} Predictions from the MLP, ResNet, and Sinter-PiNDiff models are shown from left to right. The upper and lower rows show density and grain-size evolution, respectively. Open circles denote testing measurements, solid curves show the five-model ensemble means, and shaded bands indicate $\pm3$ ensemble standard deviations.}
    \label{fig:mgo_model_comparison}
\end{figure}

\Cref{fig:mgo_model_comparison} shows that the MLP and ResNet underestimate the later density measurements at $1600\,^{\circ}\mathrm{C}$. The MLP also develops a nonmonotonic segment after its initial rise, whereas the ResNet approaches an excessively low plateau with a broad ensemble band. Sinter-PiNDiff more closely follows the rapid increase and subsequent slowing of densification, although it still underpredicts several early-to-intermediate measurements. For grain size, the MLP produces changes in slope and an extended interval of weak growth that do not follow the measured trend, while the ResNet underpredicts much of the history and exhibits substantial disagreement among ensemble members. Sinter-PiNDiff yields a smoother growth trajectory that better follows the overall experimental evolution, with increasing uncertainty at long hold times. Taken together, the density and grain-size predictions demonstrate that Sinter-PiNDiff provides more physically consistent temporal evolution and more reliable generalization to the held-out $1600\,^{\circ}\mathrm{C}$ condition than either data-driven baseline. This advantage is also reflected in its substantially lower density NRMSE across training initializations ($14.6\pm3.5$\%, compared with $39.6\pm26.0$\% for the ResNet), indicating that embedding the sintering kinetics improves predictive accuracy and reduces sensitivity to initialization under the matched evaluation protocol.

\FloatBarrier
\subsubsection{Al-doped ZnO}

\begin{figure}[ht]
    \centering
    \includegraphics[width=0.9\textwidth]{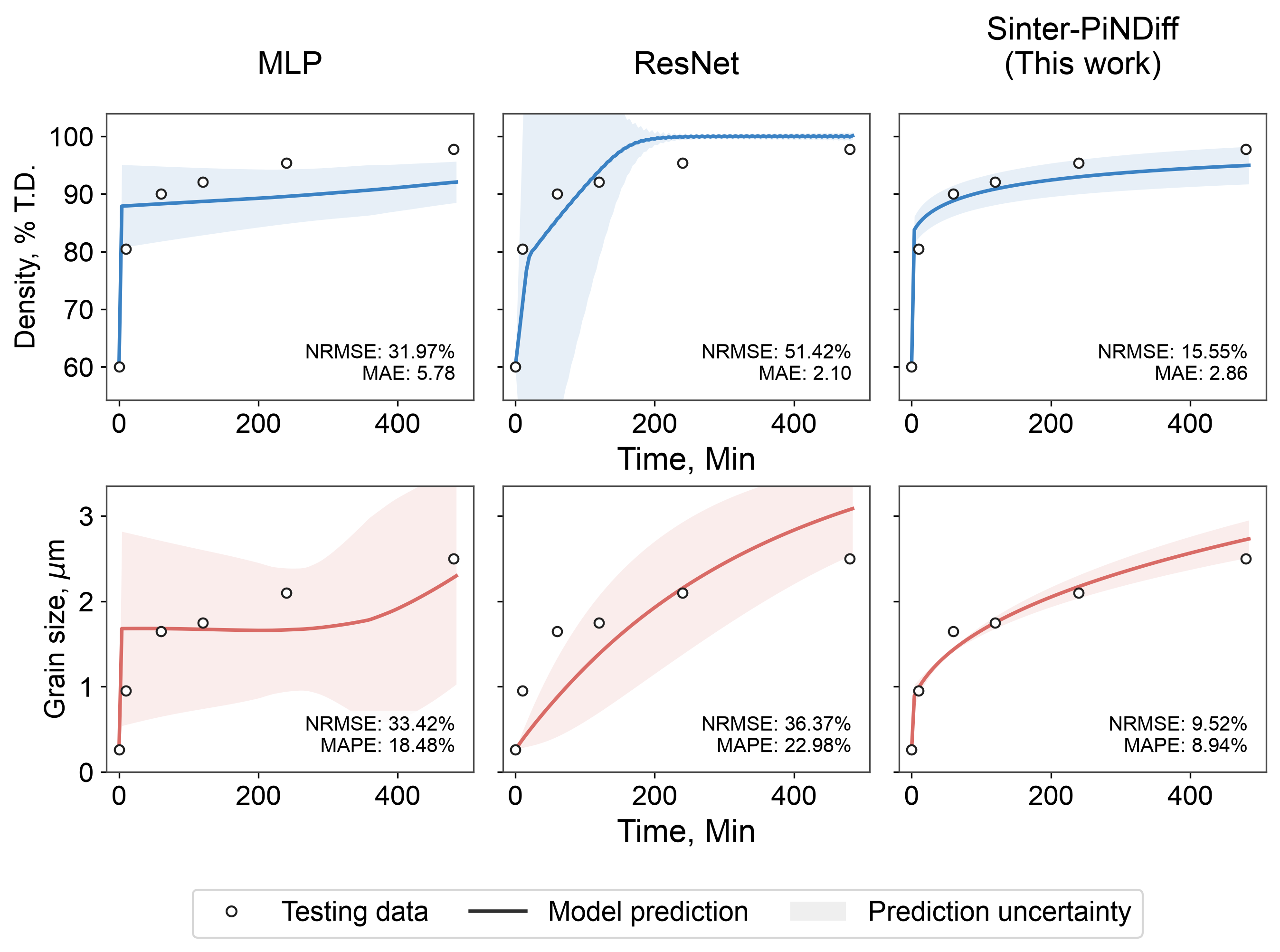}
    \caption{\textbf{Comparison of temporal predictions for $0.60$ mol\% Al-doped ZnO at $1200\,^{\circ}\mathrm{C}$.} Predictions from the MLP, ResNet, and Sinter-PiNDiff models are shown from left to right. The upper and lower rows show density and grain-size evolution, respectively. Open circles denote testing measurements, solid curves show the five-model ensemble means, and shaded bands indicate $\pm3$ ensemble standard deviations.}
    \label{fig:zno_model_comparison}
\end{figure}

\Cref{fig:zno_model_comparison} compares predictions for the withheld $0.60$~mol\% Al composition at $1200\,^{\circ}\mathrm{C}$. The MLP approximates the initial density increase as an abrupt jump followed by weak evolution and similarly produces a prolonged near-plateau in grain size. The ResNet gives smoother trajectories, but its density continues to rise beyond $100$\% theoretical density at long times, and its grain-size ensemble remains relatively dispersed. Sinter-PiNDiff captures the gradual reduction in densification rate and the continued, slowing grain growth more closely, with a particularly narrow grain-size band. Its late-time density is still somewhat below the measurements, so the improvement does not imply exact agreement at every time. Together with the reductions in both trajectory errors and both endpoint errors in \cref{tab:pindiff_baseline_comparison}, this case supports the value of retaining sintering-specific evolution laws even for composition interpolation, where the test condition lies within the training-composition range.

\FloatBarrier
\subsubsection{\texorpdfstring{CaO-doped ThO$_2$}{CaO-doped ThO2}}

\begin{figure}[htbp]
    \centering
    \includegraphics[width=0.9\textwidth]{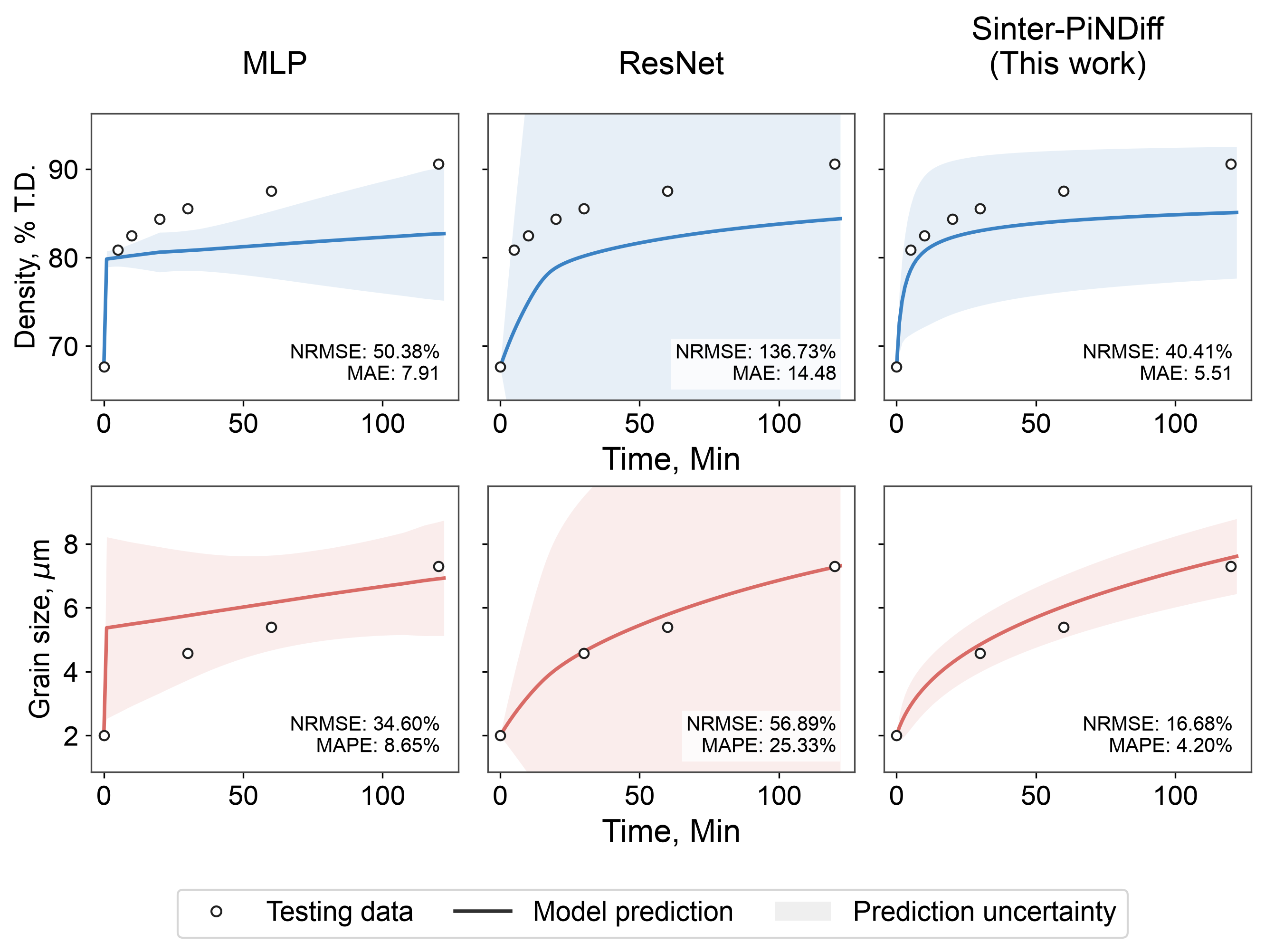}
    \caption{\textbf{Comparison of temporal predictions for $8.8$ mol\% CaO-doped ThO$_2$ at $1550\,^{\circ}\mathrm{C}$.} Predictions from the MLP, ResNet, and Sinter-PiNDiff models are shown from left to right. The upper and lower rows show density and grain-size evolution, respectively. Open circles denote testing measurements, solid curves show the five-model ensemble means, and shaded bands indicate $\pm3$ ensemble standard deviations.}
    \label{fig:tho2_model_comparison}
\end{figure}

\Cref{fig:tho2_model_comparison} considers $8.8$~mol\% CaO at $1550\,^{\circ}\mathrm{C}$, combining temperature interpolation with composition extrapolation. All three ensemble means underpredict the later density measurements, and their density predictions are much closer to one another than in the MgO comparison. For grain size, the MLP exhibits an abrupt initial increase that overestimates the early measurements, while the ResNet follows the observations with a broad ensemble band. Sinter-PiNDiff provides a gradual, concave-down growth history with appreciably less spread than the ResNet, although its mean lies above some intermediate measurements. Thus, the figure reveals an advantage in the representation and consistency of grain-growth dynamics while retaining the common density-extrapolation limitation.

The full-test comparison in \cref{tab:pindiff_baseline_comparison} provides the broader evidence of improved accuracy: Sinter-PiNDiff reduces density NRMSE from $20.1$\% for the MLP and $35.0$\% for the ResNet to $14.4$\%, and grain-size NRMSE from $27.4$\% and $31.6$\% to $19.3$\%. Its final-density MAE of $3.33$ percentage points and final-grain-size MAPE of $9.5$\% are also lower than both baselines. These aggregate gains coexist with the visible bias at the selected extrapolation condition. Across the three systems, the table establishes consistent improvement in mean predictive error, while the trajectory figures identify the specific temporal behaviors that are improved and the regimes that remain difficult.

%%%%%%%%%%%%%%%%%%%%%%%%%%%%%%%%%%%%%%%%%%%%%%%%%%%%%%%%%%%%%%%%%%%%%%%%%%%%%%%%%%%%%%%%%%%%%%%%%%%%%%%%%%%%%%%%%%%%%%%%%%%%%%%%%%%%%%%%
%%%%%%%%%%%%%%%%%%%%%%%%%%%%%%%%%%%%%%%%%%%%%%%%%%%%%%%%%%%%%%%%%%%%%%%%%%%%%%%%%%%%%%%%%%%%%%%%%%%%%%%%%%%%%%%%%%%%%%%%%%%%%%%%%%%%%%%%
%%%%%%%%%%%%%%%%%%%%%%%%%%%%%%%%%%%%%%%%%%%%%%%%%%%%%%%%%%%%%%%%%%%%%%%%%%%%%%%%%%%%%%%%%%%%%%%%%%%%%%%%%%%%%%%%%%%%%%%%%%%%%%%%%%%%%%%%

\subsection{Ablation of density feedback}
\label{sec:density-feedback-ablation-results}

\begin{figure}[ht]
    \centering
    \includegraphics[width=\textwidth]
    {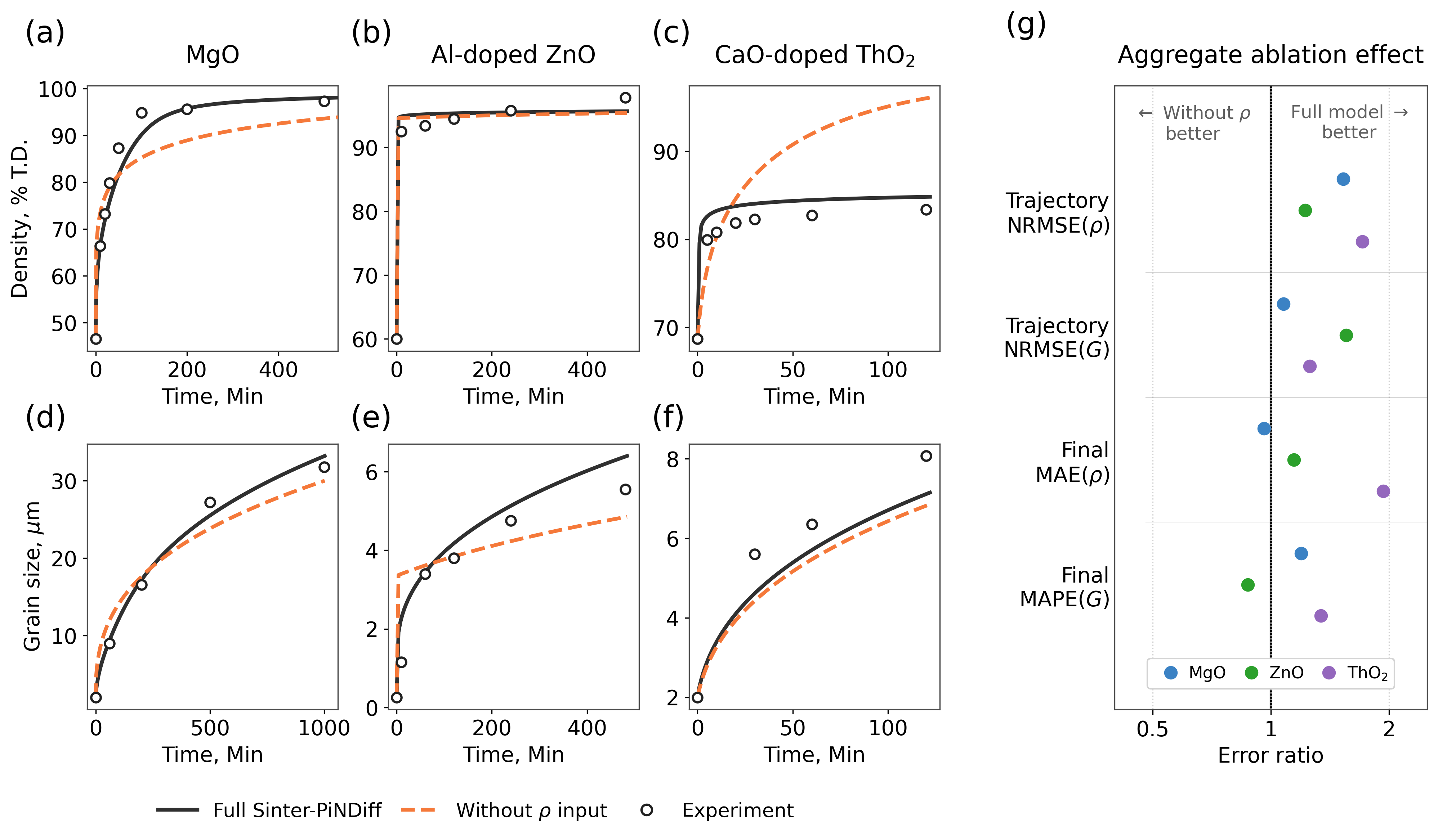}
    \caption{\textbf{Ablation of density feedback in the learned kinetic closures.}
    \textbf{(a--c)} Density and \textbf{(d--f)} grain-size predictions for representative testing conditions. Panels \textbf{(a,d)} show MgO at $1500\,^{\circ}\mathrm{C}$, panels \textbf{(b,e)} show $0.08$~mol\% Al-doped ZnO at $1200\,^{\circ}\mathrm{C}$, and panels \textbf{(c,f)} show $2.31$~mol\% CaO-doped ThO$_2$ at $1550\,^{\circ}\mathrm{C}$. Within each material system, the representative condition was selected as the testing condition whose density NRMSE ratio was closest to the median. Solid black curves denote the full Sinter-PiNDiff model, dashed orange curves denote the ablated model in which $\rho_n$ was removed from both neural-network inputs, and open circles denote experimental measurements. \textbf{(g)} Aggregate ablation error ratios over the full testing dataset for density and grain-size trajectory NRMSE, final-density MAE, and final-grain-size MAPE. The ratio is defined as the error of the model without the density input divided by the error of the full model; values greater than one therefore favor the full model, whereas values below one favor the ablated model.}
    \label{fig:density-feedback-ablation}
\end{figure}

\Cref{fig:density-feedback-ablation} evaluates the contribution of supplying the evolving density to both learned kinetic closures. Within each material system, the representative testing condition was selected as the condition whose density NRMSE ratio was closest to the median. The resulting cases are MgO at $1500\,^{\circ}\mathrm{C}$, $0.08$~mol\% Al-doped ZnO at $1200\,^{\circ}\mathrm{C}$, and $2.31$~mol\% CaO-doped ThO$_2$ at $1550\,^{\circ}\mathrm{C}$. Removing $\rho_n$ from the neural-network inputs produces the clearest changes in the later portions of the trajectories, where the evolving compact state increasingly affects the effective densification and grain-growth kinetics. For MgO at $1500\,^{\circ}\mathrm{C}$, the ablated model underpredicts the late-time density and grain size. For $0.08$~mol\% Al-doped ZnO at $1200\,^{\circ}\mathrm{C}$, the two models give similar density trajectories, but the full model more closely reproduces the continued grain growth. For $2.31$~mol\% CaO-doped ThO$_2$ at $1550\,^{\circ}\mathrm{C}$, the model without density feedback substantially overpredicts the late-time density, whereas the full model approaches the experimentally observed plateau more closely.

The aggregate error ratios in \cref{fig:density-feedback-ablation}g show that including density feedback reduces both density and grain-size trajectory NRMSE for all three material systems. The full model has lower error in ten of the twelve material--metric ablation comparisons. The two exceptions are the final-density MAE for MgO and the final-grain-size MAPE for Al-doped ZnO, for which the ablated model gives slightly lower error. Density feedback therefore does not improve every endpoint metric uniformly, but its consistent reduction of both density and grain-size trajectory errors across all three systems, together with its improvement in most endpoint comparisons, supports the use of state-dependent kinetic closures.

%%%%%%%%%%%%%%%%%%%%%%%%%%%%%%%%%%%%%%%%%%%%%%%%%%%%%%%%%%%%%%%%%%%%%%%%%%%%%%%%%%%%%%%%%%%%%%%%%%%%%%%%%%%%%%%%%%%%%%%%%%%%%%%%%%%%%%%%
%%%%%%%%%%%%%%%%%%%%%%%%%%%%%%%%%%%%%%%%%%%%%%%%%%%%%%%%%%%%%%%%%%%%%%%%%%%%%%%%%%%%%%%%%%%%%%%%%%%%%%%%%%%%%%%%%%%%%%%%%%%%%%%%%%%%%%%%
%%%%%%%%%%%%%%%%%%%%%%%%%%%%%%%%%%%%%%%%%%%%%%%%%%%%%%%%%%%%%%%%%%%%%%%%%%%%%%%%%%%%%%%%%%%%%%%%%%%%%%%%%%%%%%%%%%%%%%%%%%%%%%%%%%%%%%%%

\section{Discussion}
\label{sec:discussion}

The central result is that a common physics-integrated formulation can be independently retrained to predict coupled sintering behavior in three chemically distinct oxide systems using sparse literature measurements. MgO primarily tests temperature dependence with incomplete grain-size supervision, Al-doped ZnO adds strong composition dependence, and CaO-doped ThO$_2$ tests both interpolation and extrapolation across a largely withheld temperature--composition design. Applying the same evolution equations, neural-network architecture, and training procedure across these systems supports framework-level applicability beyond a single material-specific fit. The result extends the PiNDiff strategy previously developed for composite curing and isothermal chemical vapor infiltration to coupled density and grain-size evolution in solid-state sintering~\cite{AKHARE_curing,Akhare2024}. This retrainability applies to the model structure and learning procedure; each material system requires its own training data and independently learned network weights. Direct transfer or fine-tuning of weights learned for one material to a previously unseen oxide remains an important direction for future investigation.

The learned kinetic closures were trained only on density and grain-size histories; however, the resulting material-specific trends can be evaluated against experimental observations of sintering behavior reported in the literature. For MgO, the predicted rapid initial densification followed by diminishing density gain while grain growth continues reproduces the behavior reported by Gupta at both training and withheld temperatures~\cite{Gupta1971}. This separation between densification and coarsening is consistent with later pressureless-sintering experiments that identified a kinetic window in which grain growth could be suppressed while high density was retained~\cite{Li2015MgO}. For Al-doped ZnO, Sinter-PiNDiff reproduces the experimentally observed suppression of grain growth with increasing Al concentration and the more complex temperature dependence of densification. Han et al.\ characterized the corresponding microstructures using X-ray diffraction and electron microscopy, reporting an Al solubility limit of approximately 600 atomic ppm at $1200^{\circ}\mathrm{C}$ and ZnAl$_2$O$_4$ particles in specimens containing $0.32$--$1.20$~mol\% Al~\cite{Han2001}. These particles were found to pin ZnO grain boundaries and reduce the driving force for densification. Although phase identity and particle distributions are not provided to Sinter-PiNDiff, the predicted grain-growth suppression captures the aggregate kinetic signature associated with these experimentally observed second-phase effects.

A related interpretation applies to CaO-doped ThO$_2$. Laha and Das attributed the decrease in apparent activation energy with increasing CaO concentration to the influence of charge-compensating oxygen vacancies on cation migration~\cite{Laha1971}. Independent final-stage sintering experiments further showed that CaO can inhibit discontinuous grain growth and promote densification by maintaining vacancy transport from pores to grain boundaries; an optimum near $2$~mol\% CaO was reported, with second-phase inclusions proposed to impede grain-boundary motion above the solid-solubility limit~\cite{Jorgensen1970ThO2}. The $2.31$~mol\% composition in the present dataset lies near this experimentally identified regime, and the ability of Sinter-PiNDiff to interpolate its density and grain-size histories indicates that the learned closures reproduce much of the associated composition dependence. Conversely, the premature density plateau predicted at $8.80$~mol\% CaO indicates that extrapolation from training compositions no higher than $4.25$~mol\% does not fully reproduce the high-CaO regime. This discrepancy could reflect nonlinear changes in vacancy concentration, defect association, or second-phase behavior, but the available density and grain-size measurements cannot distinguish among these explanations.

The model comparisons help explain the value of separating a common process structure from material-dependent kinetics. An MLP must infer the complete time dependence directly, whereas a generic residual model must learn both the state increments and their dependence on the evolving state. Sinter-PiNDiff instead learns effective kinetic coefficients within prescribed rate equations. The explicit grain-size dependence constrains how the rates evolve during a rollout and links grain growth to densification, reducing the range of temporal behavior that must be inferred solely from sparse observations. The lower mean errors obtained by Sinter-PiNDiff in all twelve material--metric comparisons are consistent with a beneficial physical inductive bias under the reported training protocol. This interpretation is also supported by the abrupt or nonmonotonic MLP histories and the large ResNet ensemble spread observed in several comparison cases.

The density-feedback ablation provides additional evidence for the value of coupling the learned kinetics to the evolving material state. Removing the current density from both neural-network inputs increased both density and grain-size trajectory NRMSE in all three material systems, and the full model achieved lower error in ten of the twelve material--metric ablation comparisons. The two exceptions were the final-density MAE for MgO and the final-grain-size MAPE for Al-doped ZnO, showing that density feedback does not improve every endpoint metric uniformly. Nevertheless, its consistent benefit for both density and grain-size trajectory prediction indicates that the learned coefficients must respond to the changing compact state to reproduce the coupled temporal evolution across the investigated conditions. Under isothermal conditions, removing $\rho_n$ from the inputs makes $D_{\mathrm{nn}}$ and $k_{\mathrm{nn}}$ constant for a prescribed composition and temperature; the rates can then evolve only through the retained grain-size dependence and the prescribed density-saturation factor. The ablation therefore supports the joint density dependence of the two learned closures, although it does not isolate the individual contribution of density to $D_{\mathrm{nn}}$ and $k_{\mathrm{nn}}$. Separate single-network ablations would be required to make that distinction.

The uncertainty results are most informative when interpreted jointly with observation coverage and prediction bias. At a fixed temperature in the ThO$_2$ system, extrapolation to $8.8$~mol\% CaO generally produces wider bands than interpolation to $2.31$~mol\%. For MgO, the grain-size band broadens beyond the temperature range with direct grain-size supervision even though density remains within its supervised temperature range. These trends are consistent with weaker constraints on the learned closures during extrapolation. At the same time, the ZnO results demonstrate that sparse or strongly varying responses can produce broad bands within an interpolation domain or even at a training composition. The empirical coverage results reinforce this limitation: the $\pm3\sigma$ bands contained $63.2$--$90.5$\% of the density observations and $54.5$--$80.0$\% of the grain-size observations across the three systems. Ensemble spread therefore provides a local measure of disagreement among fitted models rather than a guaranteed measure of distance from the training data or prediction accuracy. The reported bands are descriptive ensemble intervals, not calibrated confidence or prediction intervals with a prescribed coverage probability. They exclude a separately modeled contribution from experimental and digitization noise, and all five members can share a common prediction bias.

The effective closures require a restrained physical interpretation. Agreement with density and grain-size measurements does not uniquely identify a microscopic diffusion coefficient, grain-boundary mobility, dopant mechanism, or secondary-phase distribution. Multiple closures can reproduce sparsely sampled trajectories, and this ambiguity becomes more pronounced where only density is measured. The learned coefficients should therefore be interpreted as effective kinetic quantities that absorb unresolved contributions from lattice and grain-boundary diffusion, solute segregation, secondary-phase formation, and pore--grain-boundary interactions. Density affects grain growth through the dependence of $k_{\mathrm{nn}}$ on $\rho_n$, while grain size affects densification through the explicit $G^{-m}$ dependence. The dependence of $D_{\mathrm{nn}}$ on $\rho_n$ additionally allows the effective densification kinetics to evolve with the compact state. Together, these terms produce bidirectional but reduced-order coupling without explicitly resolving pore topology, phase fractions, or grain-boundary state.

The prescribed density-saturation factor,
$\psi(\rho)=\tanh[(1-\rho)/0.05]$, provides an additional physical constraint that is distinct from the stress-intensification contribution absorbed into $D_{\mathrm{nn}}$. The factor remains close to unity at lower densities and smoothly approaches zero as the relative density approaches one, thereby attenuating the densification rate near full theoretical density without substantially altering the earlier portion of the trajectory. It should not be interpreted as a microscopic stress-intensification or transport mechanism; rather, it imposes the known limiting behavior that densification must cease at full density. Because the state is advanced using finite explicit Euler steps, the factor acts as a soft saturation constraint rather than a strict algebraic bound. Symmetric ensemble bands can also extend outside the physically admissible density range even when the ensemble-mean trajectory remains bounded. Strict enforcement would require a bounded state transformation or an explicit projection of the updated density.

The present validation remains limited to the reported isothermal experiments, their prescribed initial states, and the available sampling times. It does not establish performance for nonisothermal schedules, pressure-assisted sintering, different powder histories, or long-time behavior beyond the observed domain. The underpredicted high-CaO density histories further show that retraining the framework does not guarantee accurate extrapolation when the testing conditions involve kinetic regimes that are insufficiently represented in the training data. Additional phase and microstructural measurements would be required to distinguish insufficient training coverage from limitations of the retained rate equations and to determine whether additional state variables or stronger coupling are warranted. Because NRMSE pools the available observations, densely sampled histories contribute more strongly than sparse endpoint conditions; the separate final-state metrics therefore remain necessary for evaluating the achieved density and grain size.

These findings suggest a practical role for Sinter-PiNDiff in experimental planning. The framework cannot directly identify solute segregation, quantify secondary-phase fractions, or determine microscopic diffusion mechanisms because these quantities are not included in the model state. Instead, composition-dependent changes in the predicted density and grain-size trajectories can identify conditions where additional characterization would be most informative. For example, X-ray diffraction or electron microscopy could be concentrated near the $0.08$--$0.32$~mol\% Al range, where ZnAl$_2$O$_4$ becomes experimentally observable, and at the extrapolated $8.80$~mol\% CaO condition, where systematic prediction bias indicates a possible change in effective kinetics. The differentiable trajectories could also be used to compare the instantaneous density gain with the accompanying grain growth, $\dot{\rho}/\dot{G}$, thereby identifying temperature--time regions that provide efficient densification with limited coarsening. Such model-guided process windows would remain hypotheses until prospectively validated, but they provide experimentally testable guidance from otherwise sparse measurements.

Overall, Sinter-PiNDiff provides a practical basis for extending sparse sintering datasets across processing conditions through material-specific retraining. A natural next step is to use the ensemble predictions to identify conditions where additional density, grain-size, or phase measurements could reduce uncertainty, with particular attention to extrapolated compositions and conditions with incomplete microstructural observations. Such selection should consider residual bias as well as ensemble width because agreement among models does not ensure accuracy. Future work should also investigate transfer learning between material systems and prediction under nonisothermal conditions. The differentiable formulation further permits temperature--time optimization under competing density and grain-size objectives. These extensions require prospective validation but build directly on the demonstrated cross-material retrainability of the framework.

%%%%%%%%%%%%%%%%%%%%%%%%%%%%%%%%%%%%%%%%%%%%%%%%%%%%%%%%%%%%%%%%%%%%%%%%%%%%%%%%%%%%%%%%%%%%%%%%%%%%%%%%%%%%%%%%%%%%%%%%%%%%%%%%%%%%%%%%
%%%%%%%%%%%%%%%%%%%%%%%%%%%%%%%%%%%%%%%%%%%%%%%%%%%%%%%%%%%%%%%%%%%%%%%%%%%%%%%%%%%%%%%%%%%%%%%%%%%%%%%%%%%%%%%%%%%%%%%%%%%%%%%%%%%%%%%%
%%%%%%%%%%%%%%%%%%%%%%%%%%%%%%%%%%%%%%%%%%%%%%%%%%%%%%%%%%%%%%%%%%%%%%%%%%%%%%%%%%%%%%%%%%%%%%%%%%%%%%%%%%%%%%%%%%%%%%%%%%%%%%%%%%%%%%%%

\section{Conclusion}
\label{sec:conclusion}

Sinter-PiNDiff provides a retrainable physics-integrated framework for predicting coupled densification and grain growth from sparse experimental data. By embedding material-specific neural kinetic closures within a common set of sintering evolution equations, the framework was independently retrained for MgO, Al-doped ZnO, and CaO-doped ThO$_2$ using the same architecture and training procedure. Evaluation on withheld temperature and composition conditions demonstrated accurate interpolation and useful extrapolative capability despite incomplete density and grain-size supervision. This cross-material applicability is established at the framework level, with separate network weights learned for each material system.

Sinter-PiNDiff achieved the lowest mean error in all twelve material--metric comparisons with the MLP and ResNet baselines. For MgO, Al-doped ZnO, and CaO-doped ThO$_2$, respectively, the density NRMSE values were $14.6\pm3.5$\%, $10.8\pm1.4$\%, and $14.4\pm3.9$\%, while the corresponding grain-size NRMSE values were $8.6\pm0.7$\%, $12.1\pm4.1$\%, and $19.3\pm5.2$\%. Final-density MAEs were $2.80$, $1.56$, and $3.33$ percentage points, and final-grain-size MAPEs were $4.7$\%, $13.4$\%, and $9.5$\%. The physics-integrated formulation also produced smoother and more physically consistent temporal evolution than the data-driven baselines. The prescribed density-saturation factor further attenuated the densification rate as the relative density approached unity, providing a smooth final-stage constraint within the differentiable time integration.

The density-feedback ablation further demonstrated the value of coupling the learned kinetics to the evolving material state. Removing the current density from both neural-network inputs increased both density and grain-size trajectory NRMSE in all three material systems, and the full model achieved lower error in ten of the twelve material--metric ablation comparisons. Although the ablated model produced slightly lower errors for two endpoint comparisons, the consistent improvement in both predicted trajectories supports the joint dependence of the learned densification and grain-growth coefficients on the evolving density.

The ensemble spread generally increased in regimes with weaker experimental constraints, including composition extrapolation in CaO-doped ThO$_2$ and MgO grain growth beyond the directly supervised temperature range. Nevertheless, the persistent density bias at high CaO content shows that ensemble disagreement does not capture all sources of model error. Further validation across additional materials and processing schedules, together with calibrated uncertainty quantification, strictly bounded state formulations, and prospective experimental validation, will be required to establish broader cross-material applicability. The present results provide a foundation for using differentiable sintering models in uncertainty-informed experimental design and temperature--time optimization.

%%%%%%%%%%%%%%%%%%%%%%%%%%%%%%%%%%%%%%%%%%%%%%%%%%%%%%%%%%%%%%%%%%%%%%%%%%%%%%%%%%%%%%%%%%%%%%%%%%%%%%%%%%%%%%%%%%%%%%%%%%%%%%%%%%%%%%%%
%%%%%%%%%%%%%%%%%%%%%%%%%%%%%%%%%%%%%%%%%%%%%%%%%%%%%%%%%%%%%%%%%%%%%%%%%%%%%%%%%%%%%%%%%%%%%%%%%%%%%%%%%%%%%%%%%%%%%%%%%%%%%%%%%%%%%%%%

\section*{Data availability}
The experimental density and grain-size data used in this study were obtained from three previously published studies: Gupta, “Sintering of MgO: Densification and grain growth” (1971), for MgO (https://doi.org/10.1007/BF00550287); Han et al., “Densification and grain growth of Al-doped ZnO” (2001), for Al-doped ZnO (https://doi.org/10.1557/JMR.2001.0069); and Laha and Das, “Isothermal grain growth and sintering in pure ThO$_2$ and ThO$_2$–CaO compositions” (1971), for pure and CaO-doped ThO$_2$ (https://doi.org/10.1016/0022-3115(71)90147-4). These sources correspond to Refs. 43, 19, and 44, respectively. The processed datasets used for model training and testing and the numerical data underlying the reported results are available upon reasonable request.

\section*{Code availability}
Code is available upon request.

\section*{Acknowledgments}
The authors would like to acknowledge the funds from the U.S. Department of Energy’s (DOE) National Nuclear Security Administration (NNSA, Grants \# DE-NA0004256) in supporting this study.

\section*{Author Contributions}
Z.C., A.A., K.M., and T.L. contributed to the ideation and design of the research; Z.C., K.M., and T.L. performed the research (implemented the model, conducted numerical experiments, analyzed the data, contributed materials/analysis tools); Z.C. wrote the manuscript; A.A., K.M., and T.L. contributed to manuscript editing.

\section*{Competing interests}
The authors declare no competing interests.

\noindent\textbf{Corresponding author:} Tengfei Luo (\url{tluo@nd.edu}).

\clearpage
\bibliographystyle{elsarticle-num}
\bibliography{ref}
\end{document}